\documentclass[a4paper,fleqn,usenatbib]{mnras}

\usepackage{newtxtext,newtxmath}

\usepackage[T1]{fontenc}
\usepackage{ae,aecompl}
\usepackage{times}

\usepackage{graphicx}	
\usepackage{amsmath}	

\usepackage{acronym}
\usepackage{ulem}

\newcommand{\msun}{\mathrm{M}_\odot}
\newcommand{\rsun}{\mathrm{R}_\odot}
\newcommand{\lsun}{\mathrm{L}_\odot}

\acrodef{EoS}{equation of state}
\acrodef{SN}{supernova}
\acrodefplural{SN}[SNe]{supernovae}
\acrodef{CCSN}{core-collapse supernova}
\acrodefplural{CCSN}[CCSNe]{core-collapse supernovae}
\acrodef{RSG}{red supergiant}
\acrodefplural{RSG}[RSGs]{red supergiants}
\acrodef{SNR}{supernova remnant}
\acrodefplural{SNR}[SNRs]{supernova remnants}

\title[Origin of SESN without companions]{Formation pathway for lonely stripped-envelope supernova progenitors: implications for Cassiopeia A}

\author[R. Hirai et al.]{
Ryosuke Hirai$^{1,2}$\thanks{E-mail: ryosuke.hirai@monash.edu},
Toshiki Sato$^{3,4,5}$,
Philipp Podsiadlowski$^{6,7}$,
\newauthor Alejandro Vigna-G\'omez$^{8}$,
Ilya Mandel$^{1,2,9}$
\\
$^{1}$OzGrav: The Australian Research Council Centre of Excellence for Gravitational Wave Discovery, Clayton, VIC 3800, Australia\\
$^{2}$School of Physics and Astronomy, Monash University, VIC 3800, Australia\\
$^{3}$RIKEN, 2-1 Hirosawa, Wako, Saitama 351-0198, Japan\\
$^{4}$NASA, Goddard Space Flight Center, 8800 Greenbelt Road, Greenbelt, MD 20771, USA\\
$^{5}$Department of Physics, University of Maryland Baltimore County, 1000 Hilltop Circle, Baltimore, MD 21250, USA\\
$^{6}$Department of Physics, University of Oxford, Keble Rd, Oxford OX1 3RH, United Kingdom\\
$^{7}$Argelander-Institut f\"ur Astronomie der Universit\"at Bonn, Auf dem H\"ugel 71, D-53121 Bonn, Germany\\
$^{8}$DARK, Niels Bohr Institute, University of Copenhagen, Blegdamsvej 17, 2100, Copenhagen, Denmark\\
$^{9}$ Birmingham Institute for Gravitational Wave Astronomy and School of Physics and Astronomy, University of Birmingham,  B15 2TT, Birmingham, UK
}

\date{Accepted XXX. Received YYY; in original form ZZZ}

\pubyear{2020}

\begin{document}
\label{firstpage}
\pagerange{\pageref{firstpage}--\pageref{lastpage}}
\maketitle

\begin{abstract}
 We explore a new scenario for producing stripped-envelope supernova progenitors. In our scenario, the stripped-envelope supernova is the second supernova of the binary, in which the envelope of the secondary was removed during its red supergiant phase by the impact of the first supernova. Through 2D hydrodynamical simulations, we find that $\sim$50--90~$\%$ of the envelope can be unbound as long as the pre-supernova orbital separation is $\lesssim5$ times the stellar radius. Recombination energy plays a significant role in the unbinding, especially for relatively high mass systems ($\gtrsim18~\msun$). We predict that more than half of the unbound mass should be distributed as a one-sided shell at about $\sim$10--100~pc away from the second supernova site. We discuss possible applications to known supernova remnants such as Cassiopeia A, RX~J1713.7-3946, G11.2-0.3, and find promising agreements. The predicted rate is $\sim$0.35--1$\%$ of the core-collapse population. This new scenario could be a major channel for the subclass of stripped-envelope or type IIL supernovae that lack companion detections like Cassiopeia A.
\end{abstract}




\section{Introduction}\label{sec:introduction}
Stripped-envelope \acp{SN} are a subtype of core-collapse \acp{SN} that originate from stars that have lost most or all of their hydrogen envelope prior to the explosion. They are usually classified as type Ib, type Ic, or type IIb depending on their spectra. It has long been debated whether the progenitors of stripped-envelope \acp{SN} have lost their envelopes because of their own stellar winds \citep[e.g.][]{heg03}, or through binary interactions involving mass transfer and possibly common-envelope evolution \citep[e.g.][]{pod92}. Analysis of the environments of type Ib/c \acp{SN} supports the former scenario, in which very massive stars ($\gtrsim30~\msun$) have lost their envelopes through stellar winds \citep[]{mau18}. On the other hand, typical ejecta mass estimates of stripped-envelope \acp{SN} show very low values ($\sim2$--$4~\msun$), indicating that the progenitors originated from lower-mass stars ($\lesssim20~\msun$) that have lost their envelopes through binary interaction \citep[]{lym16,tad18,pre19}.

The most direct way to distinguish between these scenarios is to search for a surviving binary companion after the \ac{SN}. Such searches have been successful for type IIb \acp{SN}, where putative surviving companions were discovered, e.g. SN1993J \citep[]{mau04}, SN2011dh \citep[]{fol14,mau19}, SN2001ig \citep[]{ryd18}. There is also evidence of a surviving companion to a type Ibn SN2006jc \citep[]{mau16,sun20}, which is a rare subclass of type Ib SNe. These discoveries are critical indicators that the progenitors originated from binary systems. However, it is not uncommon for later observations to question whether these putative companions are true companions or line-of-sight contaminants \citep[]{fox14,mau15}.

There are also a number of cases where a companion has not been discovered despite extremely deep searches, e.g. iPTF13bvn \citep[]{fol16}, SN1994I \citep[]{van16}. This does not immediately mean that the progenitors for these stars were single massive stars. In particular, iPTF13bvn has a corresponding pre-SN image of its progenitor\footnote{It is the only type Ib SN known to date with a pre-SN progenitor detection.}. Through modelling both the pre-SN stellar properties and the rapidly declining light curve, it is firmly believed that it should have had a binary origin \citep[]{ber14,eld15}. It could be possible that the companion was a compact object, but the expected rate of such cases is extremely low \citep[]{RH17a,RH17b,zap17}.

There are also Galactic \acp{SNR} that are considered to originate from stripped-envelope SNe but have stringent upper limits on the remaining companion. One of them is Cassiopeia A (Cas A), which is known to be a type IIb \ac{SNR} from light echo spectra \citep[]{kra08}. The inferred ejecta mass is $\sim2$--$4~\msun$ \citep[]{wil03,hwa12}, which is typical of a type IIb SN and is consistent with binary progenitor models. However, the upper limits placed on any optical counterpart are so strong that no stellar companion of reasonable mass and age is allowed other than a white dwarf, neutron star or black hole \citep[]{koc18,ker19}. Another case is SNR RX~J1713.7-3946 (G347.3-0.5), which is inferred to be a type Ib/c \ac{SNR} \citep[]{kat15}. The spectral type of any remaining companion to its central compact object (1WGA J1713.4-3949) has been constrained to be later than M \citep[]{mig08}. These observational constraints seem contradictory, making it challenging to explain the origin through conventional binary evolution channels. 

\begin{figure}
 \centering
 \includegraphics[width=3.2in]{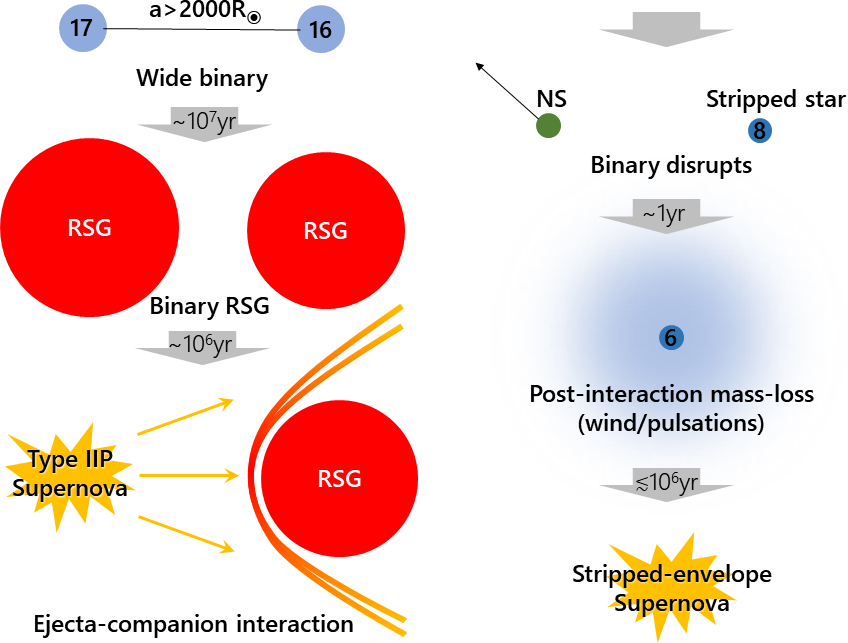}
 \caption{Schematic diagram of our fiducial scenario. Numbers on the circles express the mass of the star in solar units while numbers on grey arrows show the rough timescales between each phase. Ejecta-companion interaction and binary disruption occurs simultaneously.
 \label{fig:schematic}}
\end{figure}

In this paper we explore a new scenario that was proposed in \citet{sat20} for producing stripped-envelope \ac{SN} progenitors (Figure~\ref{fig:schematic}). In classical scenarios, the main driver of stellar mass loss are winds driven by radiative forces or mass transfer induced by gravitational forces in binary systems. Here we consider an alternative mechanically induced mass loss, which occurs through the interaction of \ac{SN} ejecta with the binary companion. This only becomes important when the companion star at the the time when the primary explodes is in its \ac{RSG} phase, where the envelope is very loosely bound. The system should be wide enough to allow the secondary \ac{RSG} to fit into its Roche lobe, implying that the system would not have experienced any previous mass-transfer events. Such systems have usually been ignored as being ``essentially single'' in binary evolution studies \citep[e.g.][]{zap17}. Although binary RSGs have not been directly observed yet \citep[]{neu20}, there is no strong reason why they should not exist. We do not rule out any possibilities that the system experienced previous mass-transfer episodes as long as the secondary is an \ac{RSG} at the point of the first \ac{SN}. However, such cases are extremely rare in binary evolution models.

At the order-of-magnitude level, the binding energies of \ac{RSG} envelopes are roughly
\begin{align}
 E_\mathrm{env} &\sim \frac{GM_\mathrm{env}M}{R}\\
&\approx6\times10^{47}\mathrm{erg}\left(\frac{M_\mathrm{env}}{10~\msun}\right)\left(\frac{M}{15~\msun}\right)\left(\frac{R}{1000~\rsun}\right)^{-1},
\end{align}
where $M$, $M_\mathrm{env}$ and $R$ are the total stellar mass, envelope mass and stellar radius, respectively. This is significantly lower than the canonical kinetic energy in \ac{SN} explosion ejecta $E_\mathrm{exp}\sim10^{51}$~erg, meaning that the ejecta can have a non-negligible effect on the companion. The fractional energy of the explosion that is intersected by the companion is 
\begin{equation}
 E_\mathrm{int}=E_\mathrm{exp}\frac{1-\sqrt{1-(R/a)^2}}{2},\label{eq:intersected_energy}
\end{equation}
where $a$ is the binary separation. We can hence estimate a maximum separation at which the intersected energy exceeds the companion's envelope binding energy $E_\mathrm{int}>E_\mathrm{env}$ as
\begin{align}
 a&<\frac{R}{\sqrt{1-\left(1-2E_\mathrm{env}/E_\mathrm{exp}\right)^2}}\\
 &\sim20~R\left(\frac{E_\mathrm{exp}}{10^{51}~\mathrm{erg}}\right)^{\frac{1}{2}}\left(\frac{E_\mathrm{env}}{6\times10^{47}~\mathrm{erg}}\right)^{-\frac{1}{2}}.\label{eq:max_sep}
\end{align}
The last expression holds as long as the binding energy is much lower than the explosion energy ($E_\mathrm{env}\ll E_\mathrm{exp}$), which is always satisfied for RSGs. Considering that RSGs can have radii of order $R\sim1000~\rsun$, this maximum separation can go up to $a_\mathrm{max}\sim100$~AU. Even beyond this separation, the ejecta can have a non-negligible impact on the envelope. However, there are other possible effects such as deflection or stellar compression during the ejecta passage \citep[]{RH18}, that can lower the efficiency of the energy transfer. So this formula only provides a rough proxy of where the ejecta--companion interaction can become important, and we resort to hydrodynamical simulations to more accurately estimate the impact of such interactions on the companion.

The collision between \ac{SN} ejecta and companion stars have been thoroughly investigated both analytically and numerically in the context of type Ia \acp{SN} \citep[e.g.][]{col70,che74,whe75,liv92,mar00,pod03,men07,kas10,pan10,pan12,liu12,liu13,mae14,nod16,bau19}. Fewer studies have been conducted for core-collapse SNe and these have mainly focused on main-sequence companions \citep[]{RH15,liu15,rim16,RH18}. In one of our previous studies \citep[]{RH14}, we have carried out hydrodynamical simulations of the impact of \ac{SN} ejecta on RSGs, which is the primary focus of this paper. The amount of mass that was removed by the impact only reached up to $\sim25\%$, which is much smaller than analogous studies for lower-mass red giants where they remove $\sim98\%$ of the envelope \citep[]{liv92,mar00}. This is partly expected since the typical binding energies of low-mass red-giant envelopes ($\sim10^{46-47}$~erg) are more than an order of magnitude lower than that of high-mass RSGs ($\sim10^{47-48}$~erg), while the explosion energies of type Ia and core-collapse SNe are similar ($\sim10^{51}$~erg). However, in \citet{RH14} we used an ideal gas law for the \ac{EoS} without the contribution of radiation pressure. In RSG envelopes, the contribution of radiation pressure is significant and this inadequate choice of \ac{EoS} can greatly alter the temperature distribution and therefore the total internal energy content of the envelope. We also only injected $10^{51}$~erg to drive the \ac{SN} without accounting for the binding energy of the envelope, meaning that the final kinetic energy of the ejecta was less than intended. We therefore expect that with a more realistic \ac{EoS} and explosion energy, the results will be notably different.

In this paper we re-examine the effect of SN ejecta colliding with RSG companions with a more realistic \ac{EoS} and explosion energy. We first outline the stellar models used in our study and the numerical method in Section~\ref{sec:method}. We then present our results in Section~\ref{sec:results} and discuss the physical process of mass removal in Section~\ref{sec:discussion}. The implications of our results, especially on the relation to the Cas A progenitor and other \acp{SN} are discussed in Section~\ref{sec:implications}. We conclude and summarize our results in Section~\ref{sec:conclusion}.

\section{Method}\label{sec:method}

\subsection{Stellar models}

The stellar models used for both the exploding and companion stars are generated using the stellar evolution code MESA \citep[v12115;][]{MESA1,MESA2,MESA3,MESA4,MESA5}. We apply sub-solar metallicity ($Z=0.0055$) for the models. This value is chosen because we have Cas A in mind as a possible target of application, which is suggested to have exploded from a sub-solar metallicity progenitor \citep[]{sat20}. However, our results are not strongly dependent on metallicity since the structure of early RSG envelopes is not sensitive to metallicity variations other than slight differences in the final masses due to the metallicity dependent wind. The exploding star models are evolved up to the core C-burning stage, after which the structure of the envelope does not change significantly up to core collapse. The companion star models are evolved up to the early core He-burning stage, which is usually $\sim10^6$~yr before core collapse. This assumes that the primary star in the binary had a lifetime $\lesssim10^6$~yr shorter than the secondary star. 

\begin{figure}
 \centering
 \includegraphics[]{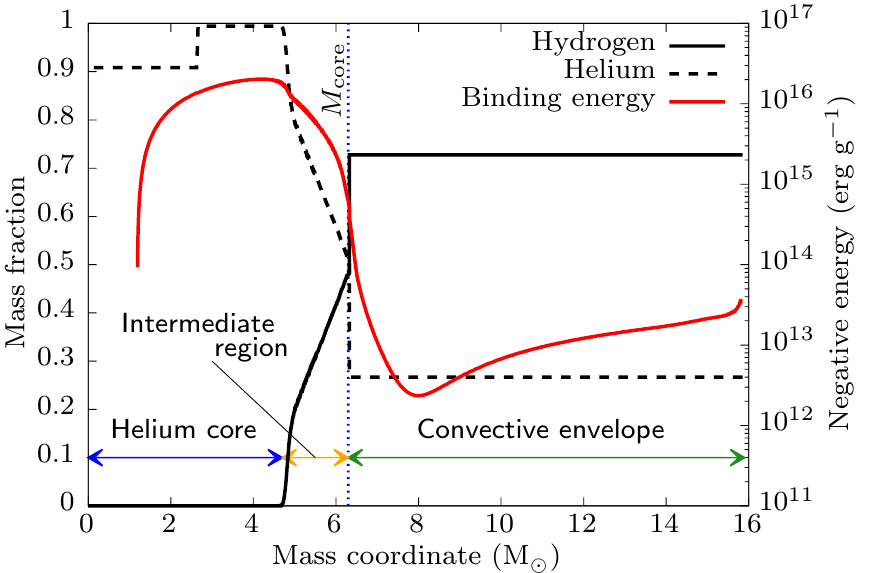}
 \caption{Chemical profile and binding energy distribution in our 16$~\msun$ early RSG model. The vertical dotted line marks the definition of our ``core'' mass. \label{fig:chemical_profile}}
\end{figure}

We display the chemical profile and binding energy distribution of one of our early core He-burning RSG models in Figure~\ref{fig:chemical_profile}. The star can be split into roughly three zones: the helium core, the convective part of the envelope and an intermediate region in between. The hydrogen fraction is flat within the outermost convective part of the envelope because it has been thoroughly mixed through convection. Then the hydrogen fraction linearly declines inwards in the radiative part of the envelope where the specific binding energy (gravitational + internal energy) is 2--3 orders of magnitude larger. This intermediate region contains the helium generated in the earlier stages of the main sequence and left over due to the contraction of the convective core. The helium core has even stronger binding energy around the edge, but also has a region of positive binding energy right at the centre. The chemical profiles of RSGs are very similar for all masses.

\begin{figure}
 \centering
 \includegraphics[]{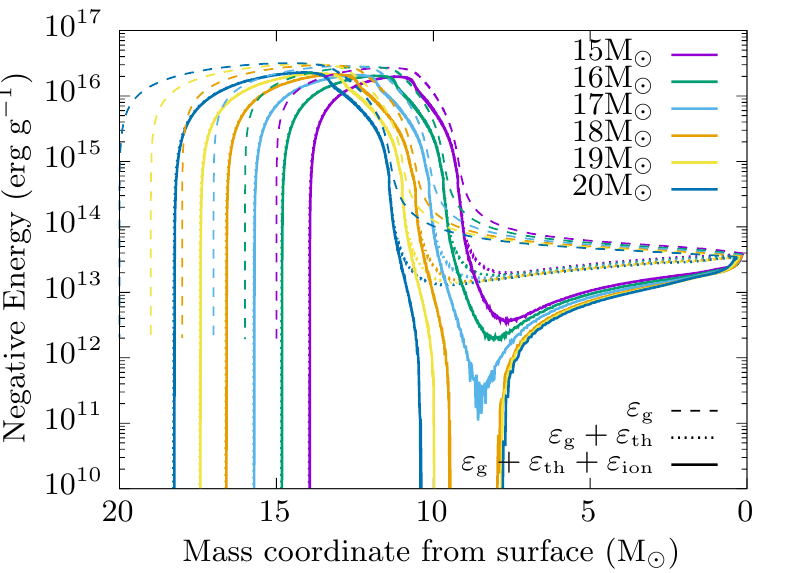}
 \caption{Specific binding energy distribution for selected RSG models. The vertical axis shows the negative local energy, and the horizontal axis is the mass coordinate from the surface. Dashed curves show only the gravitational binding energy. Dotted curves show the binding energy with the contribution of the thermal component of internal energy. Solid curves show the binding energy with thermal and ionization energy. The stellar models are taken from an early stage in the RSG phase. \label{fig:ebind_r}}
\end{figure}

The binding energy distributions for different RSG masses are summarized in Figure~\ref{fig:ebind_r}. The specific gravitational binding energy $\varepsilon_\mathrm{g}$ is almost constant throughout the envelope for all RSGs (dashed curves). For massive RSGs, radiation pressure can exceed gas pressure in the deeper layers of the envelope and hence the effective adiabatic index $\gamma$ approaches $\sim4/3$. Therefore the thermal energy can become comparable to the gravitational energy, so the combined energy approaches $\sim0$ (dotted curves). In fact in these deeper layers ($\gtrsim7.5~\msun$ from the surface), the combined binding energy becomes so small that it is comparable or even smaller than the ionization energies of hydrogen and helium. Thus if we add the contribution of ionization (a.k.a.~recombination) into the internal energy (which is what we use in Figure~\ref{fig:chemical_profile} and elsewhere unless otherwise specified), the combined binding energy can sometimes become positive (solid curves), meaning that those layers can potentially become unbound on their own if the matter on top is removed \citep[]{han94,kru16}. This effect is stronger for the higher-mass RSGs ($\gtrsim18~\msun$), so our proposed scenario favours relatively higher-mass binaries compared to lower-mass binaries. It also means that recombination energy may play an important role in ejecting the envelope, especially for higher-mass systems.

\begin{figure}
 \centering
 \includegraphics[]{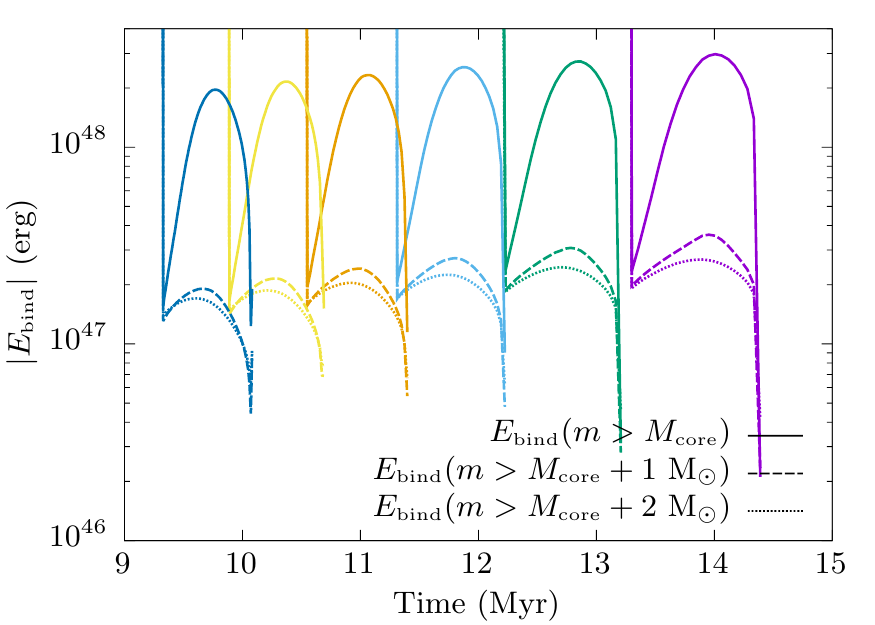}
 \caption{Time evolution of the envelope binding energies for various RSGs. Time is measured from the start of the main sequence. Colours of the curves indicate the mass of the star with the same legend as Figure~\ref{fig:ebind_r}. Binding energies were integrated from the surface of the star down to the surface of the core (solid), $1~\msun$ above the core (dashed) and $2~\msun$ above the core (dotted).\label{fig:ebind_t}}
\end{figure}

In Figure~\ref{fig:ebind_t} we show the time evolution of the envelope binding energy up to core C-burning in various RSGs with different zero-age main-sequence masses. Here we include the full internal energy in the integration of binding energy. The envelope is defined where the hydrogen mass fraction is above 0.6.  This definition distinguishes the convective part of the envelope from the rest of the star, but there is still $\sim$1--2$~\msun$ of material in the intermediate region between the hydrogen burning shell and convective envelope which is a mixture of hydrogen and helium (Figure~\ref{fig:chemical_profile}). We treat this radiative part of the envelope as part of the ``core'' because it is tightly bound and is unlikely to be affected by the \ac{SN} impact. Figure~\ref{fig:ebind_t} shows that the total binding energy of the envelope can vary by an order of magnitude during its RSG lifetime, so in principle the amount of mass unbound by an SN impact can be very sensitive to the timing of the explosion. However, the variation in the binding energy of the outer parts of the envelope ($\gtrsim1~\msun$ above the core) are much smaller during core He-burning. Thus the envelope stripping should not be too sensitive to the timing of explosion as long as it does not strip down to $<1~\msun$ above the core. The binding energy drops significantly after core helium depletion, making it even easier to strip the envelope. However, the subsequent phases are very short lived, so we do not expect many systems to have a \ac{SN} in this short time frame.

For our following hydrodynamical simulations, we choose the $16~\msun$ and $19~\msun$ RSG models as the secondary star in the binary. This allows us to study cases both with and without a positive binding energy layer in the envelope. We pair the secondary models with primary star models that have $1~\msun$ more mass than the secondary. These primary star models end their lives during the RSG phase of the companion if they are born simultaneously (see Figure~\ref{fig:ebind_t}).

\subsection{2D hydrodynamical simulations}

We carry out hydrodynamical simulations of ejecta-companion interaction following the same 2-step procedure as in our previous studies \citep[][]{RH14,RH15,RH18}. We use the hydrodynamical code HORMONE \citep[]{RH16}, which is a grid-based code that solves the hydrodynamical equations with a Godunov-type scheme. Details and some minor updates to the code are described in Appendix \ref{app:code}. 

To obtain the density distribution of the SN ejecta for our star, we first simulate the explosion of the primary star assuming spherical symmetry. The central $1.6~\msun$ is excised from the computational domain to represent the neutron star formed after SN and a large amount of energy is injected to the inner 10 grid points to artificially initiate an explosion as a ``thermal bomb''. The injected energy is set as the sum of the intended explosion energy and the absolute binding energy of the envelope. This ensures that the final kinetic energy of the ejecta has the intended value $E_\mathrm{exp}$. We choose two different explosion energies $E_\mathrm{exp}=10^{51}, 5\times10^{51}$~erg to explore how the stripped mass depends on the explosion energy. The outer boundary is set at 50 times the stellar radius and we cover the domain with 1300 grid points. The physical quantities are recorded at a fixed position outside the star to be used in the second stage.

In the second stage of the simulation, we place the companion star at the centre of a 2D axisymmetrical cylindrical grid. The symmetry axis is taken along the line connecting the centre of the two stars in the binary. Axisymmetry is a good approximation because typical SN ejecta velocities ($\sim3000$~km~s$^{-1}$) are much faster than the orbital velocities ($\sim$30--100~km~s$^{-1}$). To avoid severe numerical issues, we treat the core as a point gravitational source with a softened gravitational potential. Note that the ``core'' particle includes the radiative part of the envelope, so it contains $\sim0.5$--1$~\msun$ of hydrogen. The size of the softened region is taken to be much smaller than the stellar radius ($r_\mathrm{s}\ll R$). Details of the numerical treatment of the core are explained in Appendix \ref{app:core}. The outer boundary is set at 4 times the stellar radius except for the side where the ejecta flow in from the $+z$ direction, where we limit the boundary at 0.9 times the assumed orbital separation. The computational domain is covered by 600 grid points in the $r$ direction and 1200 grid points in the $z$ direction. We then inject the SN ejecta obtained from the first stage into the computational domain from the outer boundary. The interaction of the SN ejecta and the companion star is simulated until the bound part of the star reaches the edge of the computational domain.

The same set of simulations are carried out for various orbital separations $a$. For our lower-mass system ($M_1=17~\msun,\ M_2=16~\msun$), we choose $a=2150, 3000, 4000, 5000, 6000, 8000~\rsun$, whereas for our higher-mass system ($M_1=20~\msun,\ M_2=19~\msun$) we choose $a=3000, 4000, 5000, 6000, 8000, 10000~\rsun$. For both cases the smallest separation is chosen to be the smallest separation where the two RSGs don't exceed their Roche lobes, to ensure that there was no mass transfer in the past. For each system, we run simulations with various \acp{EoS} as listed in Table~\ref{tab:eos}. We map the pressure and density distribution from the MESA stellar models, but different internal energy distributions are assigned for different EoSs. For a given pressure and density, EoS B always yields a larger internal energy than EoS A, so comparing the results will help us to qualitatively understand the role of internal energy in the unbinding process. EoS C1--C3 stores additional components of internal energy that is only released upon recombination of certain elements. This may serve as a delayed energy source to further unbind the envelope, which is commonly discussed in the context of common-envelope evolution. See Appendix \ref{app:recombination} for details of how we implement ionization energy.

\begin{table}
 \begin{center}
  \caption{Various \acp{EoS} used in our simulations.\label{tab:eos}}
  \begin{tabular}{cl}
   \hline
   Name & Contributions\\\hline
   EoS A  & ideal gas ($\gamma=5/3$)\\
   EoS B  & ideal gas, radiation \\
   EoS C1 & ideal gas, radiation, He$^+$/He$^{++}$ ionization \\
   EoS C2 & ideal gas, radiation, H$^+$/He$^+$/He$^{++}$ ionization \\
   EoS C3 & ideal gas, radiation, H$_2$/H$^+$/He$^+$/He$^{++}$ ionization \\\hline
  \end{tabular}
 \end{center}
\end{table}

\subsection{1D follow-up simulations}

In some cases, the mass unbinding does not cease within our 2D simulations. To obtain the final asymptotic value of the unbound mass, we follow up the 2D simulations with 1D spherical simulations. We take one of the later snapshots from the 2D simulations where the companion has retained spherical symmetry, and map it onto a 1D spherical grid in HORMONE. The computational domain is extended much larger than the original 2D grid, up to 300~AU. We do not account for the material that has already overflowed the 2D computational domain, but instead place an extremely dilute atmosphere outside. This extra matter should not have any effects on the expansion of the inner material. We continue the simulation until the bound mass settles to a stationary value.

\section{Results}\label{sec:results}

\begin{figure*}
 \centering
 \includegraphics[]{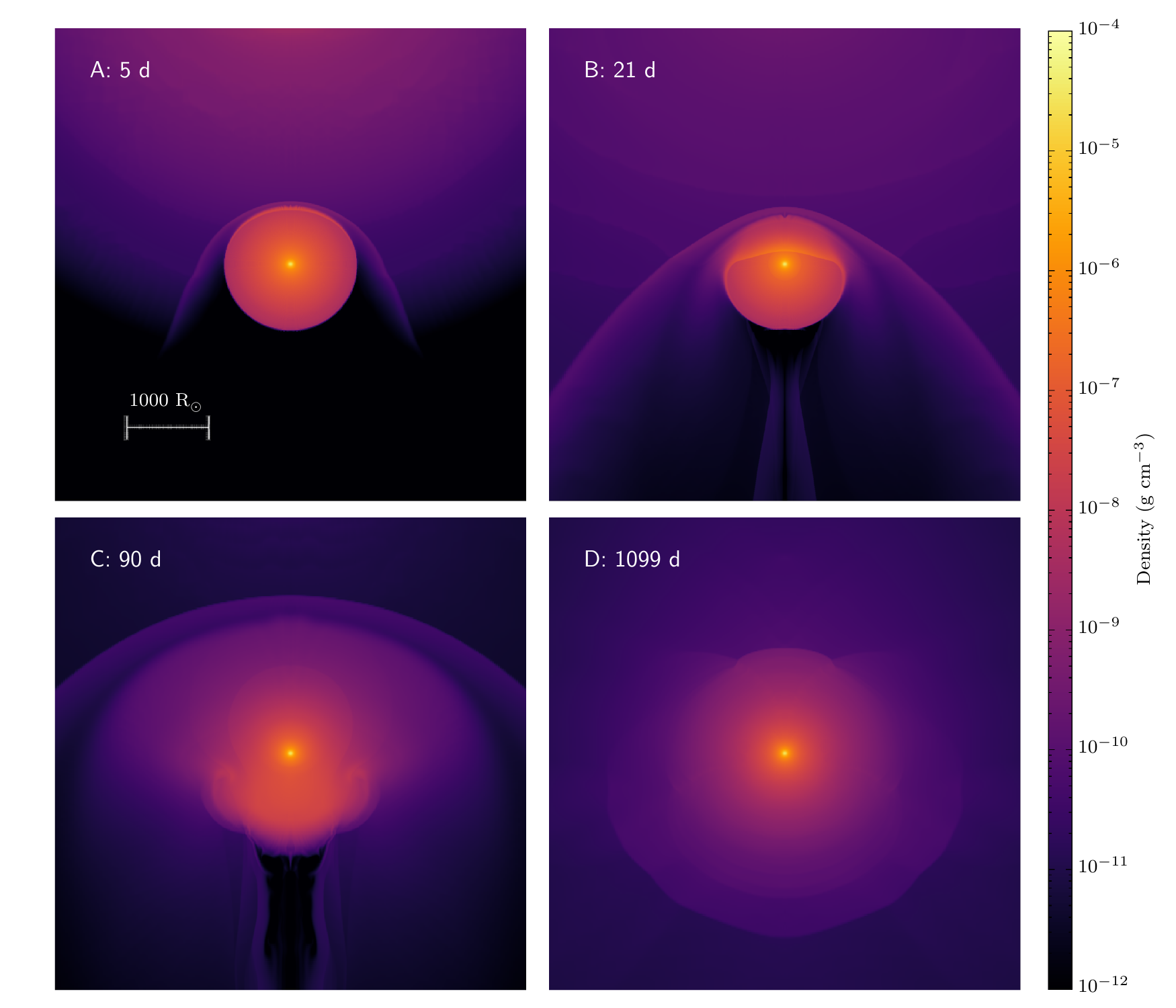}
 \caption{Density distributions of the 2D hydrodynamical simulation with $(M_1,M_2,a)=(17~\msun,16~\msun,4000~\rsun)$ and EoS C3. The explosion energy is set to $E_\mathrm{exp}=10^{51}$~erg. Each panel shows a different time snapshot. The exploding star is located above the panel.\label{fig:snapshots}}
\end{figure*}

Here we describe the dynamics of the ejecta-companion interaction of our reference model, which is a system with $(M_1, M_2, a)=(17~\msun, 16~\msun, 4000~\rsun)$ and run with EoS C3. Figure~\ref{fig:snapshots} shows various snapshots of the 2D simulation. The basic dynamics of the interaction are very similar to previous work \citep[]{mar00,RH14} and are alike among all our models. As soon as the SN ejecta reach the surface of the star, a forward shock is formed that propagates through the star and the reverse shock forms a bow shock (Panel A). The forward shock propagates faster in the outskirts of the envelope and is slower along the axis because it is decelerated as it ascends the density gradient (Panel B). We observe a small density dip at the contact discontinuity around the symmetry axis, which is a common numerical artefact of axisymmetric simulations. This feature seems to have negligible effects on the overall dynamics of the simulation. As the shock sweeps through, the envelope is heated and quickly starts to expand. Once the shock traverses the central core, it then accelerates as it runs down the density gradient on the other side of the star. The shock that travelled along the surface meets with the central shock on the other side of the star and violently pushes off a chunk of material as it penetrates the surface (Panel C). After that, the whole star expands spherically due to the heat excess (Panel D). The outer material closely follow a homologous expansion while in some models, the inner parts turn around and start falling back onto the bound star.

The unbound mass is usually calculated by integrating the mass over the cells that have positive total energy $\frac{1}{2}\mathbfit{v}^2+\varepsilon_\mathrm{int}+\phi>0$, where $\mathbfit{v}$ is the velocity, $\varepsilon_\mathrm{int}$ the specific internal energy and $\phi$ the gravitational potential of that cell \citep[e.g.][]{RH18}. We shall call this the ``energy criterion''. In the past we have used an alternate criterion that we called the ``Bernoulli criterion'', where we integrate the cells which have positive Bernoulli constants $\frac{1}{2}\mathbfit{v}^2+\varepsilon_\mathrm{int}+p/\rho+\phi>0$, where $p$ is pressure and $\rho$ is density \citep[]{RH14}. This is based on the Bernoulli theorem in which the Bernoulli constant is conserved along streamlines in stationary flows. The two criteria are based on slightly different assumptions and show quite different values during the shock-sweeping phase. However, they eventually converge to the same stationary value after a sufficiently long time. The benefit of the Bernoulli criterion was that it converges to the final value at a much earlier time \citep[]{RH14}, but we now find that this is not always the case, especially for severely stripped cases.

Defining the unbound mass in the runs with ionization energy is not trivial. In the energy criterion, it implicitly assumes that the internal energy will eventually be fully used to accelerate the material to escape velocity. This assumption is valid as long as the material is optically thick while the internal energy does work to accelerate the matter, which is a good approximation for the marginally unbound regions (the strongly unbound regions have already reached escape velocity so the internal energy term is negligible anyway). For our EoS C runs, we include ionization energy in the internal energy, and therefore the internal energy is always larger than or equal to the no-recombination case (EoS B) for any given density and temperature. Some simulations of common-envelope evolution have shown that including this extra contribution of ionization energy in the energy criterion defined above yields much larger values for the unbound mass and thus a successful ejection of the whole envelope \citep[]{nan15}. However, the ionization energy of a given mass element can never be used for acceleration unless it is released in a sufficiently optically thick region where the energy can be quickly thermalised. Otherwise the released energy is simply transported away by radiative diffusion or convection \citep[e.g.][]{sab17,iva18}. Hence the addition of ionization energy to the energy criterion involves an additional assumption that this extra energy reservoir will be fully thermalised and efficiently used for accelerating the material. This assumption is not present in the original energy criterion, making it difficult to directly compare the results. To provide a fair comparison with our no-ionization energy runs, we divide the internal energy into two parts: thermal energy $\varepsilon_\mathrm{th}$ and ionization energy $\varepsilon_\mathrm{ion}$, where $\varepsilon_\mathrm{th}+\varepsilon_\mathrm{ion}=\varepsilon_\mathrm{int}$ (see Appendix \ref{app:recombination}). By only using the thermal energy in the energy criterion ($\frac{1}{2}\mathbfit{v}^2+\varepsilon_\mathrm{th}+\phi>0$), we can directly compare the unbound masses from our simulations with and without recombination energy under the same assumptions. The results should be identical if no recombination takes place, but shall start to deviate as soon as part of the ionization energy has been released as thermal energy. We will call this the ``thermal criterion'', and this is used to compute the unbound mass in all models shown in the following sections.

\begin{figure}
 \centering
 \includegraphics[]{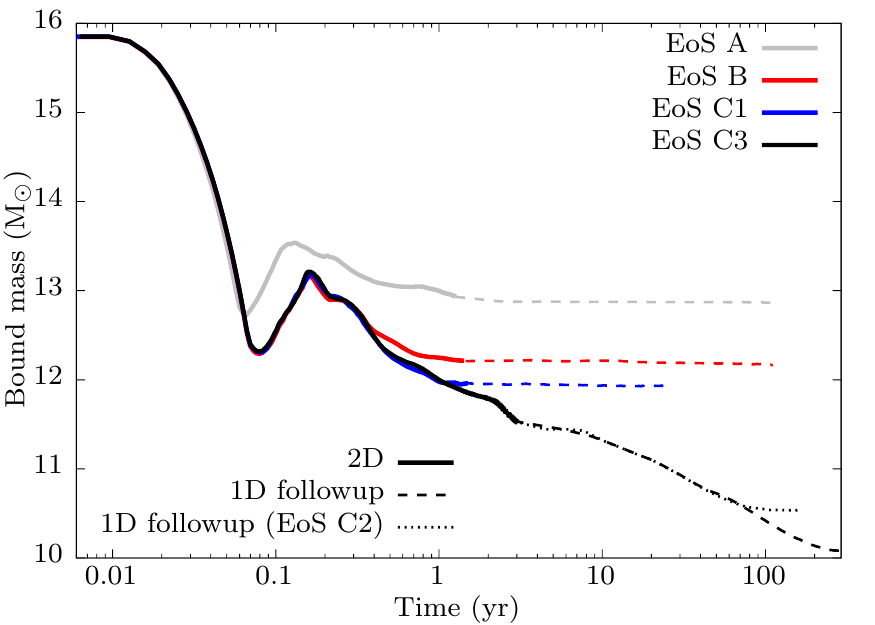}
 \caption{Time evolution of the total bound mass of the companion in the simulations for the reference model. Colours indicate results for runs with different EoSs. The solid portions of the curves show the 2D simulation results, whereas the dashed portions show results of the 1D followup simulations. The dotted curve is a 1D followup of the solid black curve but with EoS C2.\label{fig:timeevo4000}}
\end{figure}

In Figure~\ref{fig:timeevo4000} we show the time evolution of the bound mass for the reference system. Different colours are results for simulations with different EoSs. All curves initially dive down as the shock propagates up the density gradient. Then the bound masses go back up as the shock descends the density gradient on the other side. This apparent fallback is an artifact of the way the bound mass is defined, and is commonly seen in similar studies \citep[e.g.][]{pan10,RH14,RH18,liu15,rim16}. Then the bound mass goes down again and approaches a stationary value.

We immediately see that the final bound mass can differ by up to $\sim3~\msun$ depending on the applied EoS. As expected, the run with an ideal gas EoS has the least amount of unbinding because the amount of internal energy is least. The difference between the EoS A (grey) and EoS B (red) curves is $\sim0.7~\msun$, which suggests that the initial internal energy does also play a role in unbinding the envelope. However, the difference in EoS leads to different shock strengths, so the reduction in unbound mass may not be directly attributed to the initial internal energy.

The blue and black curves show results for the simulations with different species of ionization energy included in the EoS. These curves deviate from the EoS B (red) curve only after $\sim0.4$~yr as the marginally bound material expands and adiabatically cools below the helium ionization temperature. The helium recombination energy is released as a delayed energy source and unbinds more material. Then the EoS C1 (blue) and EoS C2/C3 (black) curves start to deviate at around $\sim$1~yr, where hydrogen recombination kicks in. In the EoS C1 curve we do not include hydrogen ionization energy in the EoS, so it does not generate the second wave of energy release. The EoS C3 (black dashed) curve then deviates from the EoS C2 (black dotted) curve at $\sim70$~yr as molecular hydrogen starts to form and releases more energy.

At first glance, it may seem that helium recombination only plays a minor role in the unbinding. Much more mass is removed in the hydrogen recombination stage. However, even though helium recombination alone does not provide enough energy to unbind matter, it is crucial for expanding the outer layers to the hydrogen recombination temperature. Without the push from helium recombination, the hydrogen recombination cannot kick in and therefore no unbinding happens. This multi-stage rocket effect makes it difficult to predict the role of recombination in unbinding the envelope.

There are many debates in the context of common-envelope evolution whether hydrogen recombination takes place in an optically thick enough region where the released energy is efficiently thermalised and used for acceleration \citep[]{sab17,iva18}. In our simulations, we find that the optical depth of the hydrogen recombination layer is $\sim$1--100, meaning that it is likely to thermalise the photons released from recombination. Even if the atomic hydrogen recombination energy is efficiently thermalised, it can still be transported away by radiative fluxes. On the other hand, molecular hydrogen recombination will take place in a region where the opacity is significantly lower and hence it will not be a useful source of energy.  Therefore the true unbound mass should lie somewhere between the EoS C1 and C2 curves. More detailed simulations with frequency-dependent radiation transport is required to properly evaluate how efficiently the hydrogen recombination energy can be used.

It is also possible that at some point, the temperature cools enough to start condensating dust ($T\lesssim1500$~K). The dust grains have large opacities and can be accelerated up to escape velocity by capturing radiation from the interior star \citep[]{lam99,gla18}. At sufficiently high densities, the dust grains are strongly coupled to the gas and will unbind more mass. We checked how much of the bound mass is eventually cooled down below $T<1500$~K in our simulations. For most models this is $<0.05~\msun$, and even the largest case (the $a=4000~\rsun$ model) is $<0.2~\msun$. So dust formation will not play a major role in unbinding more mass. However, the already unbound matter will all eventually cool below the dust condensation temperature, so its asymptotic coasting velocity may be affected by radiative acceleration acting on dust grains.

\begin{figure}
 \centering
 \includegraphics[]{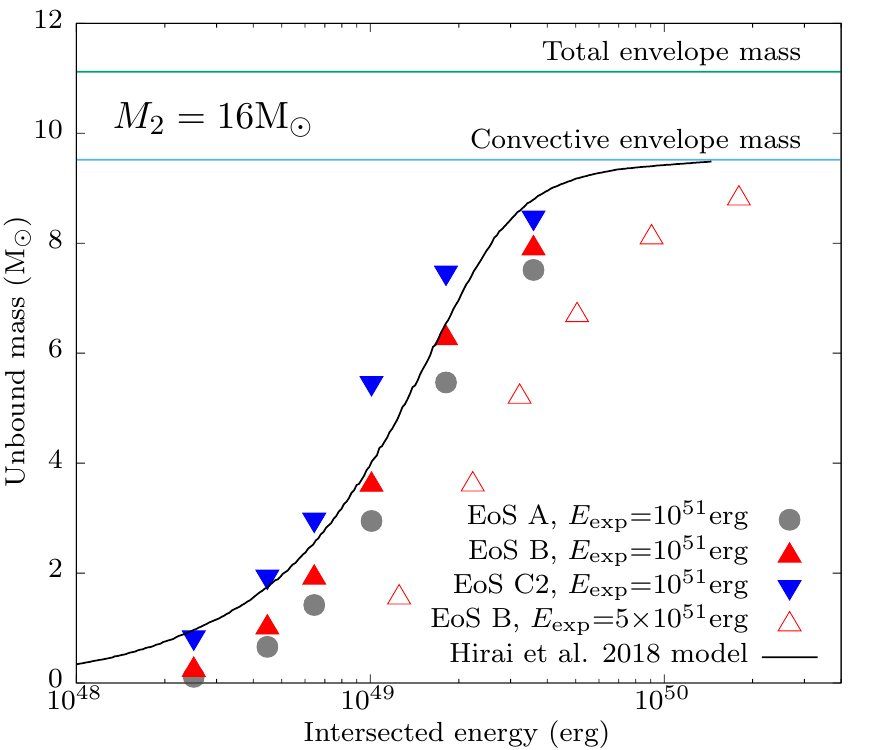}
 \caption{Final unbound mass as a function of intersected energy for all our simulations with $M_2=16~\msun$. Grey circles are results for simulations run with the ideal gas EoS. Red filled triangles were run with the ideal gas + radiation EoS, whereas the red open triangles were run with the same EoS but with $E_\mathrm{exp}=5\times10^{51}$~erg for the explosion. Simulations run with the full EoS are shown as blue triangles. The light blue line marks the mass of the convective envelope. The green line shows the total envelope mass above the helium core. The black curve is computed based on the energy injection model presented in \citet{RH18}.\label{fig:mub_energy}}
\end{figure}

Figure~\ref{fig:mub_energy} summarizes the final unbound masses obtained from our simulations for the lower-mass system ($M_1=17~\msun, M_2=16~\msun$). We plot it against intersected energy, which is a function of orbital separation and explosion energy (Eq.~(\ref{eq:intersected_energy})). The total unbound mass is largest for the model with smallest orbital separations, where $\sim8~\msun$ of the envelope has been removed through the interaction. Even for this case, more than $\sim1~\msun$ of gas from the convective envelope remains bound to the star in the simulation, justifying our approximation of replacing the inner parts with a point mass. The amount of unbound mass decreases as the separation is widened, and becomes almost negligible at $a\gtrsim8000~\rsun$. A simple estimate based on Eq.~(\ref{eq:max_sep}) yields a maximum separation of $a\sim20000~\rsun$ where the intersected energy is larger than the total binding energy of the envelope. Our results show that hardly any mass becomes unbound beyond $a>8000~\rsun$, indicating that a large fraction of the intersected energy is not used to unbind the envelope. For all systems, the unbound mass was smallest when a purely ideal gas EoS was used (grey circles) and largest when the full EoS was used (blue triangles), while the ideal gas + radiation EoS models lie in between (red triangles). We also find that the unbound masses for the $E_\mathrm{exp}=10^{51}$~erg and $E_\mathrm{exp}=5\times10^{51}$~erg models do not align with each other. This indicates that the unbound mass is not determined by the intersected energy alone, which was the basis of our simple estimates in Eq.~(\ref{eq:max_sep}).

\begin{figure}
 \centering
 \includegraphics[]{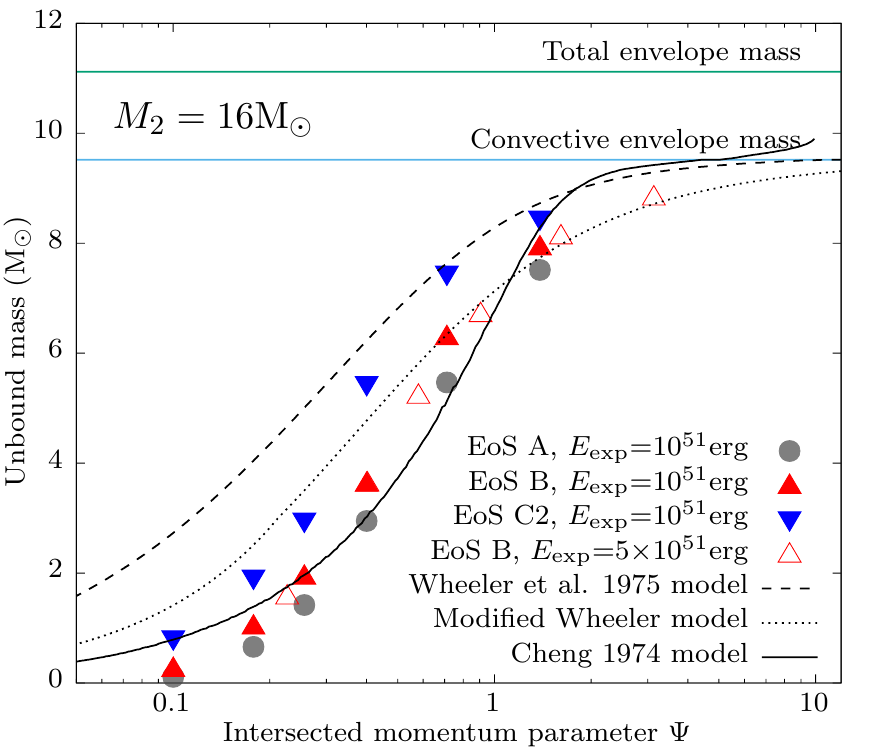}
 \caption{Final unbound mass as a function of the dimensionless parameter $\Psi$ defined in Eq.~(\ref{eq:psi}). Symbols are the same as in Figure~\ref{fig:mub_energy}. Black curves are computed based on analytic models explained in the text.\label{fig:mub_psi}}
\end{figure}

In Figure~\ref{fig:mub_psi}, we plot the unbound mass as a function of the dimensionless parameter $\Psi$, defined as
\begin{eqnarray}
 \Psi\equiv\frac{1}{4}\frac{M_\mathrm{ej}}{M_2}\frac{R^2}{a^2}\left(\frac{v_\mathrm{ej}}{v_\mathrm{es}}-1\right),\label{eq:psi}
\end{eqnarray}
where $M_\mathrm{ej}$ and $v_\mathrm{ej}$ are the SN ejecta mass and velocity, and $v_\mathrm{es}$ is the escape velocity of the RSG. It is a measure of the total momentum intersected by the secondary, first introduced in \citet{whe75}. Here the red solid and open triangles align well with each other, meaning that the unbound mass is determined by the incoming momentum, not the energy. The mass unbinding process will be discussed in more detail in Section \ref{sec:massremoval}.

\begin{figure}
 \centering
 \includegraphics[]{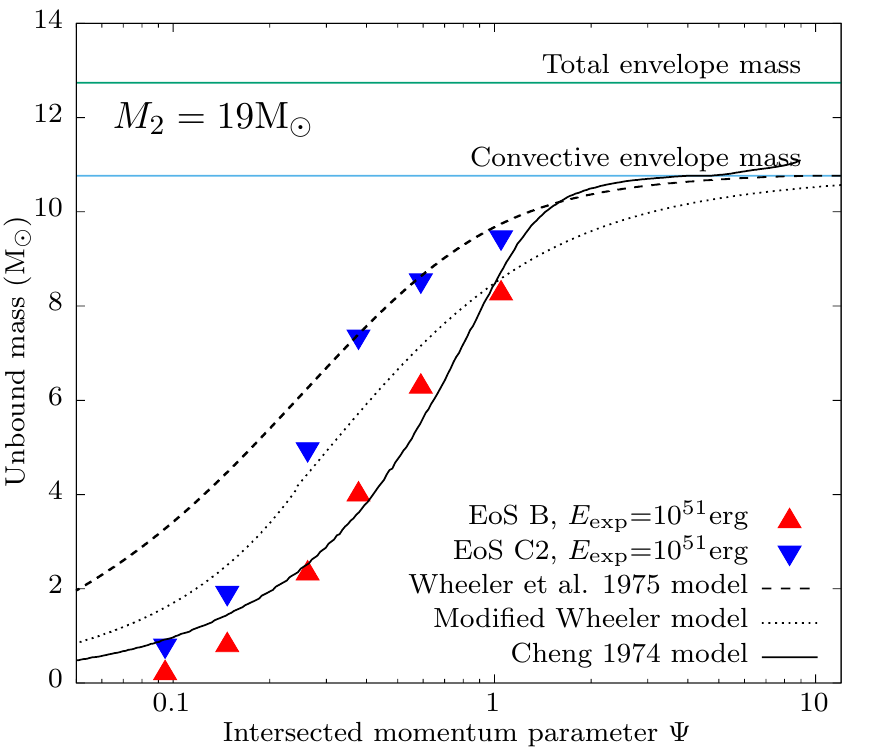}
 \caption{Same as Figure~\ref{fig:mub_psi} but for the $M_2=19~\msun$ models.\label{fig:mub19_psi}}
\end{figure}

The same results for the $M_2=19~\msun$ models are displayed in Figure~\ref{fig:mub19_psi}. The fractional unbound mass of the envelope for the EoS B runs are very similar to the $M_2=16~\msun$ models due to the similar stellar structure. The role of recombination in the $M_2=19~\msun$ models is larger than for the $M_2=16~\msun$ models, as expected from Figure~\ref{fig:ebind_r}.  For example, in the $a=5000~\rsun$ model ($\Psi\sim0.4$), $\sim4~\msun$ is unbound through the direct dynamical interaction (red triangles) but an additional $\sim3~\msun$ is unbound due to recombination (blue triangles). This extra unbinding significantly boosts the possibility for the secondary to become a stripped-envelope SN progenitor.

Although our results are not dependent on metallicity, there may be a weak dependence on the helium fraction. First, it affects the mean molecular weight which feeds into the EoS, so it will affect the thermal energy distribution of the envelope. Second, it will have an effect on the ionization energy budget. The total ionization energy is only weakly dependent on the helium fraction, but it directly affects the ratio of helium and hydrogen recombination. Because helium and hydrogen recombination occur at different temperatures and therefore different times, changing the ratio can alter the dynamics of the outflow. However, we believe that this will only have a minor effect on the final unbound masses unless the helium fraction is unusually high or low.

\section{Discussion}\label{sec:discussion}
\subsection{Mass removal process}\label{sec:massremoval}

The mass removed through the ejecta-companion interaction in our simulations falls short of our simple estimate in Section~\ref{sec:introduction}. Here we attempt to understand the reason for this and compare with simple models that predict the amount of unbound mass. 

In Figure~\ref{fig:mub_energy} we overplot a model curve (black solid) based on the semi-analytic model presented in \citet{RH18}, which explains the amount of injected energy and its distribution for main-sequence companions. In the model, the total injected energy is only a small fraction of the intersected energy. More than $>3/4$ of the energy is thermalised at the bow shock and is not used to heat the companion. Another half is simply deflected away, so the combined injection efficiency is 
\begin{equation}
 \frac{E_\mathrm{inj}}{E_\mathrm{int}}=\frac{1}{2}\frac{\gamma-1}{\gamma+1}.
\end{equation}
The injected energy is then distributed in the following form
\begin{equation}
 \Delta\epsilon(r)=\frac{E_\mathrm{inj}}{m_\mathrm{heat}}\frac{\min[1,m_\mathrm{heat}/m(r)]}{1+\ln(M_2/m_\mathrm{heat})},
\end{equation}
where $m$ is the mass coordinate from the surface and $m_\mathrm{heat}$ is the efficiently heated mass, which we estimate as $m_\mathrm{heat}=M_\mathrm{ej}(1-\sqrt{1-(R_2/a)^2})/4$. This empirical function well describes the excess energy distribution for the case of main-sequence companions where almost no mass becomes unbound \citep[see Figure~18 in][]{RH18}. We here assume the injected energy distribution is the same for RSGs and simply integrate the mass elements that have positive total energy (the sum of kinetic, thermal and gravitational energies) after the interaction to determine the amount of unbound mass via the thermal criterion. The model curve is in good agreement with the red solid triangles, only slightly overestimating the unbound mass at the low and high ends. However, the model depends only on the intersected energy (it only weakly depends on the intersected mass through $m_\mathrm{heat}$), and cannot explain why the $E_\mathrm{exp}=5\times10^{51}$~erg models have much lower unbound masses for the same intersected energy.

Figures~\ref{fig:mub_psi} and \ref{fig:mub19_psi} have three model curves overplotted. The dashed curve is the unbound mass calculated from the analytic model presented in \citet{whe75} \citep[see also][]{men07}. It has two components, stripping and ablation, both estimated from momentum conservation along each radial ray of ejecta. Stripping refers to immediate mass removal due to shared momentum and ablation refers to mass removal as the matter expand due to the heat deposited by the shock. We find in \citet{RH18} that this model greatly overestimates the unbound mass for cases where the removed mass is small. Here we find consistent results in which the model greatly overpredicts the unbound mass for the low-$\Psi$ end whereas the discrepancy is smaller for the more severely stripped cases (high-$\Psi$). We apply one modification to the Wheeler model to account for the deflection of ejecta, which was proven to be important in \citet{RH18}. For each radial ray of ejecta, we assume that only $\cos^2\theta$ of the momentum is transferred to the envelope, where $\theta$ is the angle between the line connecting the centres  of the two stars and the line from the centre of the companion to the position where the ejecta impact the surface. The remaining momentum is taken away by the ejecta that are deflected tangentially to the surface of the star \citep[see Figure~13 in][]{RH18}. This modified Wheeler model is displayed as dotted curves in Figures~\ref{fig:mub_psi} and \ref{fig:mub19_psi}. Note that both of these curves do not use any information about the internal energy of the star, so should not be compared with the results from simulations with ionization energy. The modified Wheeler model is in relatively good agreement with the simulation results at the high-$\Psi$ end, suggesting that the envelope is mostly stripped by momentum in these cases. On the other hand, the envelope is primarily unbound through ablation for the low-$\Psi$ cases and therefore the Wheeler model breaks down.

The solid black curves are based on a simple model of ablation due to heat deposited by the forward shock \citep[]{che74}. We greatly simplify the problem by only considering the shock propagation along the axis connecting the two stellar centres. The shock velocity is determined by momentum conservation
\begin{equation}
 v_\mathrm{shock}(r)=\frac{\Omega M_\mathrm{ej}v_\mathrm{ej}}{\Omega M_\mathrm{ej}+m(r)},
\end{equation}
where $\Omega\equiv(1-\sqrt{1-(R/a)^2})/2$ is the solid angle of the secondary. The specific energy deposited into the envelope is $v_\mathrm{shock}^2$ in the strong shock limit, meaning that greater specific energy will be deposited near the surface, decreasing inwards as the shock decelerates. We then assume that the material will become unbound where the excess energy exceeds the local gravitational binding energy. We see that despite the many simplifications, the model is in rough agreement with the simulated results in the intermediate-$\Psi$ regime ($0.3\lesssim\Psi\lesssim1$).

The simple shock heating model omits many details. For example, we only consider the shock propagation along the symmetry axis, which is equivalent to assuming the shock propagates outside-in in a spherically symmetrical way. In the actual situation, the shock propagates from one side of the star and will have curved trajectories as it ascends the density gradient. Also, it only considers the propagation of the forward shock. In reality, a reverse shock propagates into the SN ejecta too. As the reverse shock reaches the inner edge of the ejecta it in turn sends a rarefaction wave propagating back towards the forward shock, which eventually catches up. Once it has caught up, the rarefaction wave will decelerate the forward shock and therefore the energy deposition will decrease too. 

\begin{figure}
 \centering
 \includegraphics[]{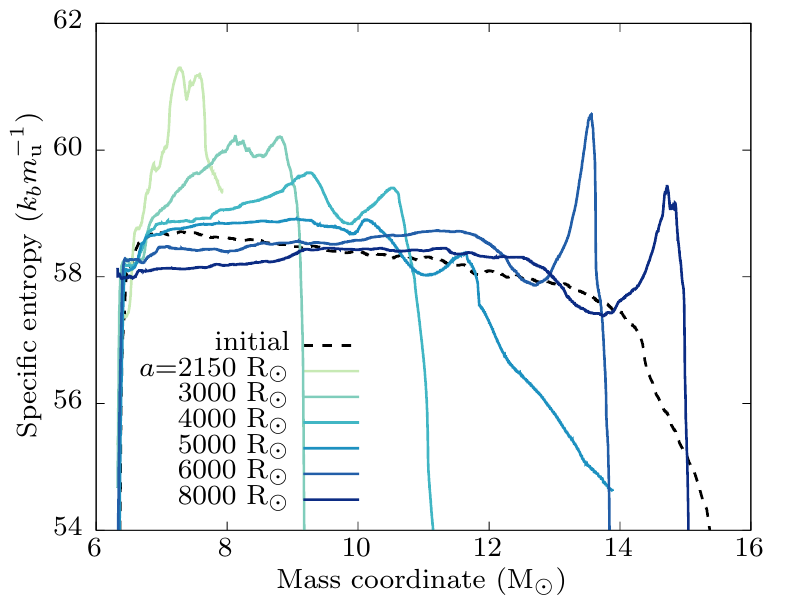}
 \caption{Spherically averaged specific entropy distributions at the final snapshot of the 2D simulations from the $M_2=16~\msun$, EoS C3 runs. The dashed curve shows the initial entropy distribution of the star.  \label{fig:entropy_1d}}
\end{figure}

\begin{figure}
 \centering
 \includegraphics[]{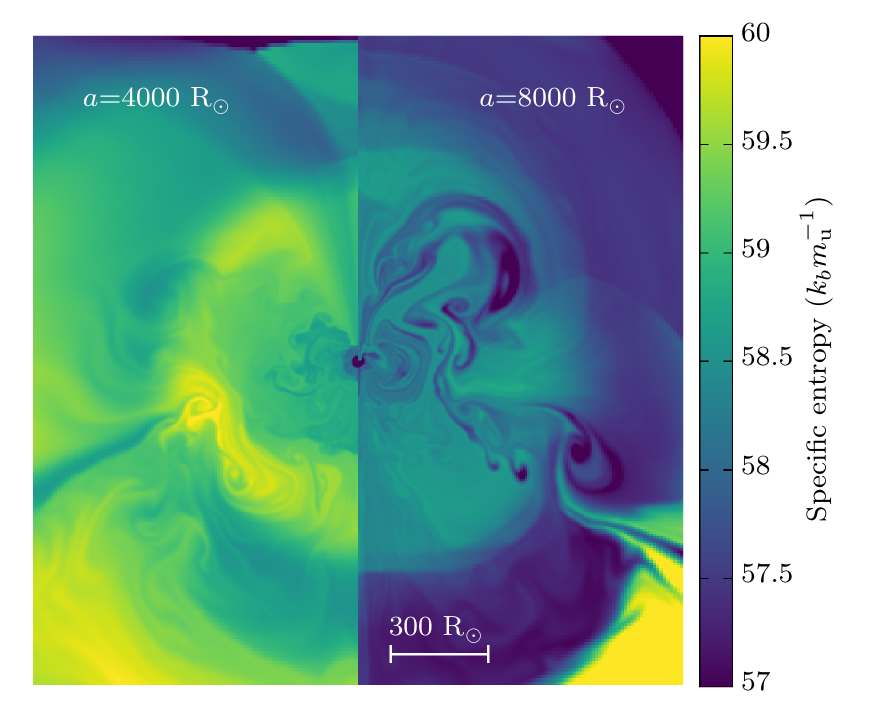}
 \caption{Distribution of entropy in our 2D simulations at a snapshot 2~yr after the SN. The left and right panels show models run with different orbital separations.\label{fig:entropy_2d}}
\end{figure}

Another assumption of the simple ablation model is that the deposited energy will eventually be converted into the kinetic energy of that mass element.  However, the energy can be redistributed through the star in the subsequent expansion phase, which can significantly impact how much mass becomes unbound. We show in Figure~\ref{fig:entropy_1d} the spherical averaged entropy distribution from the final snapshot of our 2D simulations. It is clear from the $a=2150$--$4000~\rsun$ models that the energy excess provided by the ejecta-companion interaction is larger in the outer layers and decreases as it goes deeper into the star, justifying the shock heating model. All curves have a steep drop at the outer edges because the thermal energy has already been converted to kinetic energy in these layers. We also see that in the larger separation models, the post-interaction entropy in the inner region is lower than the initial value. This is because of convective energy transport. The energy deposition by the SN is very asymmetric, and induces large-scale convective motions. Figure~\ref{fig:entropy_2d} shows the entropy distribution towards the end of the simulation, which clearly illustrates the development of such flows. Notice the low entropy plumes reaching to the centre in the right panel, whereas the convective eddies are smaller in the left panel. This convection transports part of the central heat outwards, leading to more unbinding of the envelope when the energy is transported to the marginally bound regions, which may explain why the model underpredicts the unbound mass in the intermediate separation models (Figures~\ref{fig:mub_psi} and \ref{fig:mub19_psi}). On the other hand, this process sucks out energy from the central part, which may be the reason why the model overpredicts in the high-$\Psi$ end (Figure~\ref{fig:mub_psi}). It should be noted that the convective motions may have been artificially amplified due to the axisymmetric treatment of our simulation. More realistic treatment of convection may lead to less energy transport, but it is not straightforward to predict whether this will enhance or reduce the mass loss.

To sum up, the modified Wheeler model describes the mass removal process well when the envelope is severely stripped. This implies that most of the envelope is stripped by the momentum of the SN ejecta and ablation plays a minor role. The more weakly stripped models are better described by the Cheng model, meaning that ablation is the major contributor to the mass removal. The Cheng model breaks down at the lowest impact end where the strong shock limit breaks down. Taking the smaller of the two models seems to be the best match to our results. Recombination energy induces an additional $\sim$1--3$~\msun$ of mass loss, and the contribution is larger for higher-mass systems. It does not help to unbind the last bit of envelope left on the core, but helps most in the intermediate cases when only about half of the envelope is unbound kinematically. Further investigation is required to properly understand the physical process of the mass removal after the ejecta-companion interaction.

Although the mechanism is quite different, our situation is similar to common-envelope phases in which energy is dynamically deposited into an RSG envelope. From our comparison of an ideal gas EoS case and an ideal gas + radiation EoS case, the role of internal energy in the envelope unbinding process is not dominant, nor is it negligible. This suggests that the mechanism of common-envelope evolution in massive stars may not be dramatically different from that of low mass analogues where radiation pressure can be neglected.

\subsection{Post-interaction evolution}\label{sec:post-interaction}

Our hydrodynamical simulations show that a substantial fraction of the envelope ($\lesssim0.9$) can be removed through ejecta-companion interaction. The classification of the secondary SN is ultimately determined by how much mass is lost through the interaction and its post-interaction evolution. For example, our most extreme model only leaves $\sim2~\msun$ of the envelope, which can easily be lost in the post-interaction evolution. In these cases the secondary SN would most likely be observed as a type Ib/c or IIb SN. There is more mass left in the intermediate separation models, but it is still possible that this will be mostly lost in the post-interaction evolution.

At the end of our simulation, the companion is still out of thermal equilibrium. This leads to an increase in surface luminosity while it radiates away the remaining excess energy. During this period, the wind mass-loss rates may be enhanced, leading to additional mass loss. Even without an enhanced mass-loss phase, there is enough time ($\sim10^6$~yr) before the companion explodes to lose another $\sim$1--2$~\msun$ if regular RSG wind mass-loss rates apply \citep[]{dej88,bea20}.

It is also possible that the envelope becomes dynamically unstable due to the sudden mass loss. RSG envelopes are known to develop strong dynamical pulsations when the luminosity-to-mass ratio is high \citep{heg97}. Some studies suggest that these pulsations can lead to large mass loss by driving strong winds \citep[]{yoo10}, or through pulsation-driven shocks \citep[]{cla18}. In some cases the RSG may lose the entire envelope through this process. \citet{cla18} suggests that the critical condition for this instability is roughly $\log[(L/\lsun)/(M/\msun)]\gtrsim4.1$, which can only be achieved by stars with $M\gtrsim25~\msun$ in single star models. In our scenario, the luminosity-to-mass ratio can shoot up after the interaction, abruptly sending the star into the instability regime even for stars with $M\lesssim25~\msun$. This applies particularly for the relatively higher-mass systems ($M_2\gtrsim19~\msun$), where the late-phase RSG luminosity, explosion energy of the primary SN\footnote{It is still debated whether stars in this mass range successfully explode, but the predicted explosion energies are higher if they do \citep[e.g.][]{nak15,ert16}.} and unbound mass due to ejecta-companion interaction are higher, all of which are favourable for increasing the luminosity-to-mass ratio.

Both RSG wind and dynamical pulsations have highly uncertain mass-loss rates. However, as discussed above, at least $\gtrsim$1--2$~\msun$ of mass can be lost before the secondary explodes if standard RSG mass-loss rates are applied.  This is enough for our most severely stripped models to explode as a stripped-envelope SN. Thus we propose that our new scenario can contribute to some fraction of the stripped-envelope SN population, specifically for the cases where no companion is observed. The fraction of stripped-envelope SNe produced through this channel depends on how large the post-interaction mass loss is. So by assuming this is the only channel that creates the lonely stripped-envelope SN progenitors, it may be possible to probe the post-interaction mass-loss rates by comparing the ratio of stripped-envelope SNe with and without companion detections.

\section{Implications}\label{sec:implications}

Our sets of hydrodynamical simulations clearly demonstrate that the scenario described in Figure~\ref{fig:schematic} can plausibly work under certain conditions. In this section we apply the proposed model to several observed \acp{SNR} and discuss possible observational signatures. Then we discuss how other stripped-envelope \acp{SN} or type IIL \ac{SN} progenitors can also be produced through this channel. We also give very rough estimates of the occurrence rate.

\subsection{Cassiopeia A}\label{sec:casa}
Cas A is one of the most well studied \acp{SNR}. It is known to be from a type IIb SN from light echo spectra \citep[]{kra08}. The prevailing theory for type IIb SN progenitor formation is through stable mass transfer in binary systems \citep[]{yoo17,ouc17,sra19}. Ejecta mass estimates for Cas A are $\sim2$--$4~\msun$ \citep[]{wil03,hwa12}, which translates to a helium core mass of $\sim$3--6$~\msun$ and therefore zero-age main-sequence masses of roughly $\sim$15--20~$\msun$. Recent modelling of X-ray spectra further suggest that the progenitor had a sub-solar metallicity \citep[]{sat20}, meaning that the stellar wind would have been relatively weak. All of the above strengthens the case that the progenitor was unlikely to have formed through a single star channel.

To enable stable mass transfer, the binary companion cannot be too small compared to the primary mass, so the companion should have been relatively massive too. Recent observations have attempted to find this companion that should not be too far from the centre of expansion. However, they instead place extremely strong upper limits on any remaining companion and conclude that there was no stellar companion to the progenitor upon explosion \citep[]{koc18,ker19}. Most binary evolution scenarios have been ruled out because of this, and the only remaining channels are binary mergers \citep[]{nom95}, binary disruption after mass transfer from the secondary \citep[]{zap17}, or the companion is a compact object. All require rather fine-tuned assumptions for the physics involved in order to produce the Cas A progenitor.

Our new scenario is more straightforward and can robustly produce the Cas A progenitor given the right initial conditions. For example, in our $(M_1, M_2, a)=(17~\msun, 16~\msun, 2150~\rsun)$ model, the secondary is left with only $\sim2.5~\msun$ above the core after the primary SN. If $\sim2~\msun$ is lost through the post-interaction RSG wind, it will become a $\sim6~\msun$ stripped-envelope star with only $\lesssim0.5~\msun$ of the envelope ($\lesssim0.1~\msun$ of hydrogen) left by the time it collapses. If this star explodes, it can reproduce the ejecta mass, the type IIb spectrum and the apparently single nature of Cas A.

\begin{figure}
 \centering
 \includegraphics[width=3.4in]{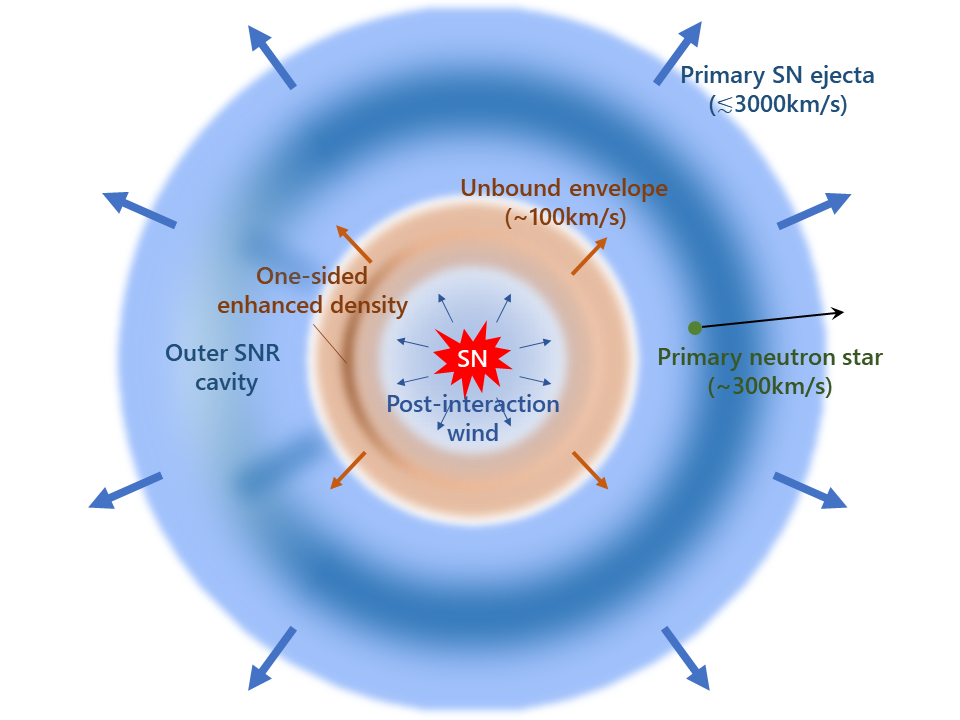}
 \caption{Illustration of the predicted circumstellar matter distribution in our scenario. The outer \ac{SNR} (blue shell), the unbound envelope (brown shell) from the secondary RSG and the primary neutron star (green circle) are all ejected in the first SN of the system. The post-interaction wind was emitted over the duration between the first and second SN. The red feature indicates the second SN. The location of the shells and the neutron star are not to scale.\label{fig:csm_structure}}
\end{figure}

If this was the case, it means that another SN occurred near the location of Cas A within the past $\lesssim10^6$~yr, which could have some interesting observational features. In Figure~\ref{fig:csm_structure} we illustrate the possible circumstellar matter distribution in this scenario. The bulk of the primary SN ejecta will be located far outside, forming an old \ac{SNR}. We expect there to be a cavity in this remnant where the companion blocked the outflow of the ejecta. Part of the blocked ejecta matter is simply deflected, so it may form a cone-shaped wall around the cavity. The size and brightness of this old \ac{SNR} will depend on the structure of the interstellar medium and the delay time between the first and secondary SNe. In some environments, it could have already dissolved into the interstellar medium. Deeper inside, the matter that was unbound from the secondary due to the ejecta-companion interaction will be coasting at much slower velocities, forming an inner shell. As we see in Figure~\ref{fig:snapshots} Panel C, a chunk of unbound material will coast in the same direction as the cavity of the outer \ac{SNR}. So there may be a significant density enhancement in this direction within the inner shell. Interior to the inner shell, the secondary star will blow out a bubble with its own wind and the second SN will occur within this environment.

The radius of the inner shell can be roughly estimated as 
\begin{equation}
 r_\mathrm{sh}\sim100~\mathrm{pc}\left(\frac{v_\mathrm{sh}}{10^2~\mathrm{km~s}^{-1}}\right)\left(\frac{\Delta t}{10^6~\mathrm{yr}}\right)\label{eq:shell_radius}
\end{equation}
where $v_\mathrm{sh}$ is the coasting velocity of the unbound envelope and $\Delta t$ the delay time between the two SNe. For the coasting velocity we choose the escape velocity of the RSG, but it can be much smaller for the marginally unbound inner parts. We choose the total RSG lifetime of the secondary as $\Delta t$ but this can also be smaller depending on the mass ratio of the binary, so we consider $r_\mathrm{sh}\sim100$~pc as a conservative upper limit. It is known that there is an H$\alpha$ cloud located 10--15~pc outside of Cas A in the North East direction \citep[]{van71,wei20}. The location of this cloud is within the ballpark of our inner shell estimate and the one-sided nature of the shell is consistent with our model. No other scenario can naturally explain the one-sided distribution of this shell. Therefore the location and structure of this H$\alpha$ cloud may be a smoking gun feature that the Cas A progenitor was indeed produced through our proposed channel. The estimated mass of the H$\alpha$ cloud is $\gtrsim1.5~\msun$, consistent with the prediction of $\lesssim$7--8$~\msun$ from our scenario\footnote{This is an upper limit because the whole envelope is not ejected one-sided. A fraction of the envelope is ablated spherically.}.

The time it takes for the secondary SN to reach the inner shell can then be estimated as
\begin{equation}
 \tau_\mathrm{int}\sim10^4\mathrm{yr}\left(\frac{v_\mathrm{sh}}{10^2~\mathrm{km~s}^{-1}}\right)\left(\frac{v_\mathrm{ej}}{10^4~\mathrm{km~s}^{-1}}\right)^{-1}\left(\frac{\Delta t}{10^6~\mathrm{yr}}\right),
\end{equation}
where $v_\mathrm{ej}$ is the ejecta velocity of the secondary SN. Here we have assumed a free expansion for the SN ejecta, which is not a bad approximation for Cas A since it has just recently entered the Sedov phase \citep[]{pat09}. Cas A is only $\sim350$~yr old \citep[]{tho01}, so despite the uncertainties, the bulk of the SN ejecta most likely have not reached the inner shell yet. It may be possible that the ejecta start overtaking the inner shell in the next $\sim10^{3-4}$~yr and emit X-rays from the interaction.

\begin{figure}
 \centering
 \includegraphics[width=3.2in]{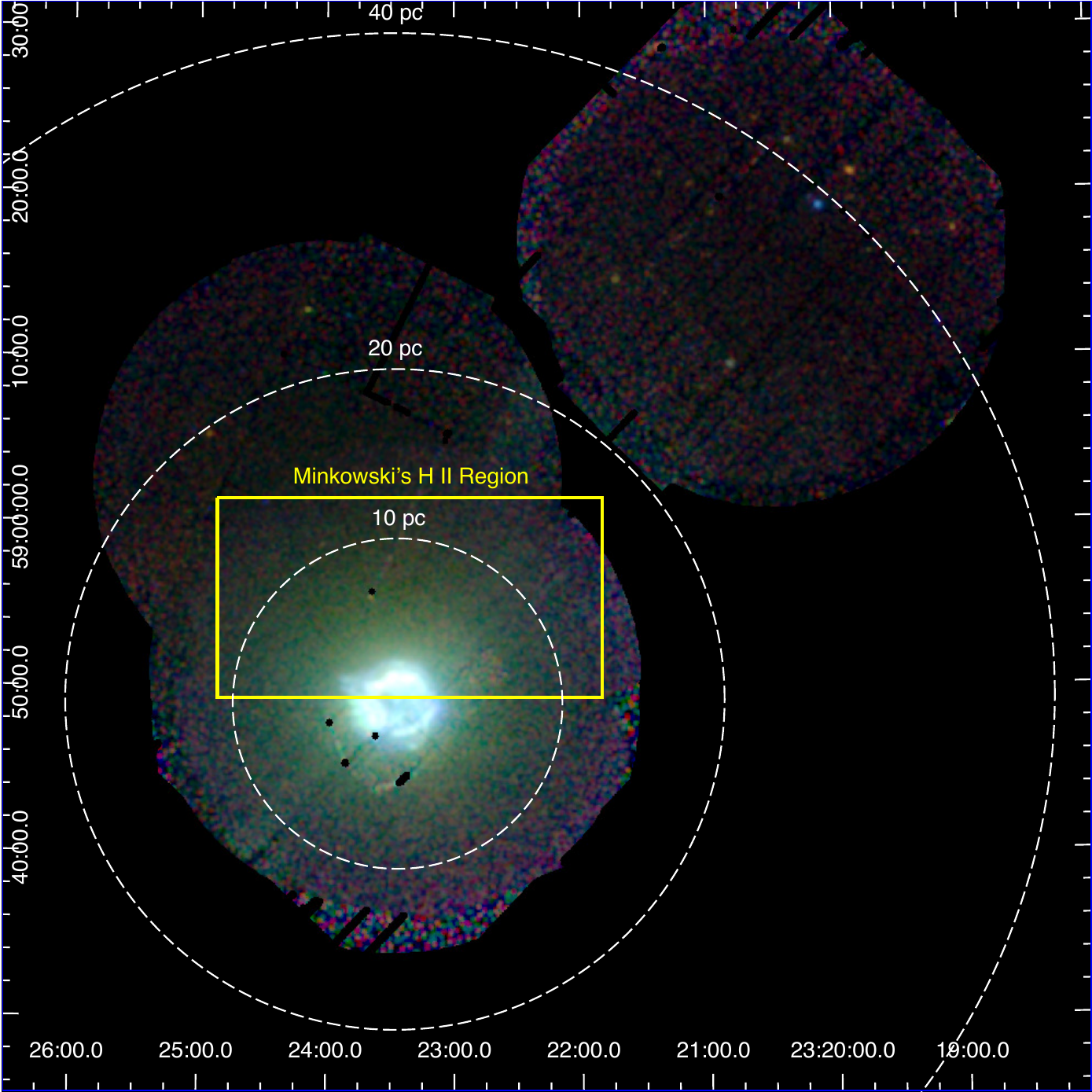}
 \caption{X-ray image (red: 0.4--0.75 keV, green: 0.75--1.3 keV, blue: 2.0--7.2 keV) around Cas A by {\it XMM-Newton} (Obs.ID 013755031, 0165560101, 0400210101, 0782961401). The dashed circles show 10, 20 and 40 pc radii from Cas A. There are no obvious X-ray features that appear to be the remnant of the primary supernova within $\lesssim$50 pc. The yellow box shows the faint H II region called ``Minkowski's H II region'' \citep[see][]{van71,wei20}. \label{fig:xray}}
\end{figure}

Detection of an outer \ac{SNR} would provide strong support for our scenario. However, current deep observations by {\it XMM-Newton} show no sign of an \ac{SNR} surrounding Cas A in X-rays within $\lesssim$50~pc (Figure~\ref{fig:xray}). At least, the presence of an SNR in the Sedov phase (younger than several $10^4$ yr) could be ruled out from this image.
Whether we can observe the remnant of the first SN, however, strongly depends on the environment. The plasma temperature of \acp{SNR} declines over time, making it more and more subject to absorption. Some relatively old ($\sim10^4$~yr) \acp{SNR} such as the Cygnus loop whose electron temperature is $\lesssim$ 0.5 keV are still bright below 1 keV due to the low column density of the environment \citep[$n_{\rm H}\sim10^{20}$~cm$^{-2}$;][]{uch09}. On the other hand, the environment of Cas A is much denser with column densities of $n_{\rm H}\gtrsim10^{22}$~cm$^{-2}$ \citep[e.g.][]{hwa12}, which makes it difficult for low-energy X-rays typical to older remnants to reach us. So the non-detection of an outer \ac{SNR} does not rule out our scenario but instead places a loose lower limit to the delay time $\Delta t\gtrsim10^4$~yr. This lower limit is consistent with the delay time inferred from the location of the H$\alpha$ cloud (Eq.~\ref{eq:shell_radius}). A future intense search for old SNR features in the vicinity of Cas A by X-ray and/or radio will be useful in testing our hypothesis.

If the binary was disrupted $\sim10^{5-6}$~yr ago, the primary neutron star should be located at a distance
\begin{eqnarray}
 r_\mathrm{NS}\sim300~\mathrm{pc}\left(\frac{v_\mathrm{kick}}{300~\mathrm{km~s}^{-1}}\right)\left(\frac{\Delta t}{10^6~\mathrm{yr}}\right),
\end{eqnarray}
away from Cas A, where $v_\mathrm{kick}$ is the kick velocity imparted to the neutron star. There are 13 known pulsars with projected distances to Cas A within 300~pc (or $\lesssim$5~deg assuming a distance of 3.4~kpc) according to the Australia Telescope National Facility (ATNF) pulsar catalogue \citep[]{man05}.

\begin{figure}
 \centering
 \includegraphics[]{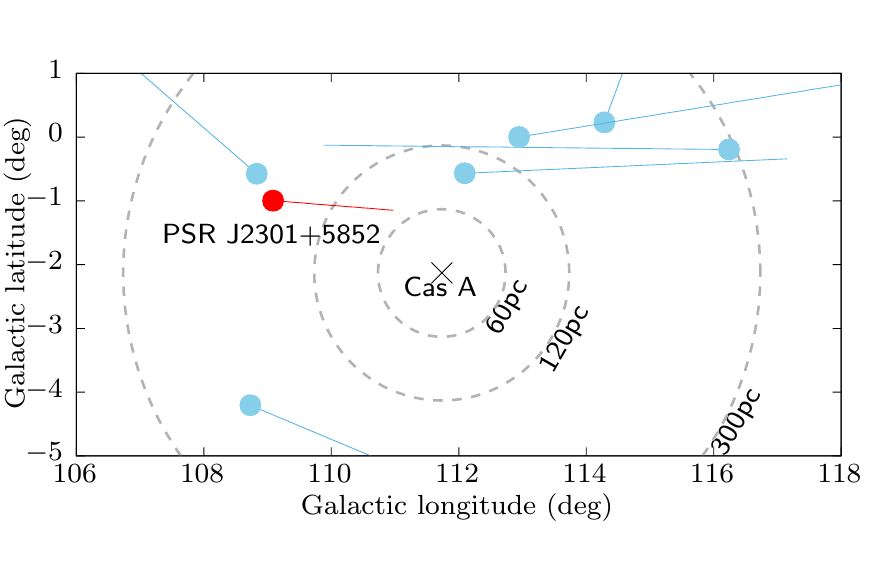}
 \caption{Location and proper motion of PSR~J2301+5852 along with several other pulsars within $\sim$5~deg from Cas A. The circles indicate the current locations of the pulsars and the lines point back in the direction of the proper motion for $10^6$~yr.\label{fig:pulsar}}
\end{figure}

Among those pulsars, we find one \citep[PSR~J2301+5852;][]{fah81} which has many properties that are consistent with being associated with Cas A. First, the characteristic age is $\sim2.35\times10^5$~yr, which is within the delay time upper limit ($\lesssim10^{6}$~yr) and consistent with the non-detection of an outer SNR ($\gtrsim10^4$~yr). Characteristic ages of pulsars are known to be poor indicators of their true ages, but having it in the right ballpark is reassuring. It is also consistent with the delay time required for having the inner shell located at $\sim$10--15~pc. Second, the proper motion is roughly pointing radially away from Cas A as illustrated in Figure~\ref{fig:pulsar} \citep[]{ten13}. It does not directly point to the centre of Cas A, but it should be noted that after the first SN, the secondary should also have a proper motion due to the disruption of the binary. Typical pre-SN orbital velocities are $\sim$30--100~km~s$^{-1}$, so it is possible that the Cas A progenitor had drifted by $\sim$30--100~pc from the first SN site. The unbound envelope shell should also have the same initial net momentum so the relative position should not change much during the drift. We see from Figure~\ref{fig:pulsar} that the projected distance from the Cas A site $\sim 10^6$~yr ago was $\sim$70~pc. Given the uncertainty in the delay time, the projected distance can be as close as $\sim$60~pc. Thirdly, the estimated distance to PSR~J2301+5852 is $\sim$3.2~kpc \citep[]{kot12}, which is in close proximity with the inferred distance to Cas A of $\sim3.4$~kpc \citep[]{ree95}. 
Although it is tempting to associate this pulsar with the first SN, PSR~J2301+5852 is an anomalous X-ray pulsar that lies within its own $\sim10^4$~yr old SNR \citep[CTB~109;][]{kot12,nak15b} so does not seem to fit into our picture. Some of the other pulsars in the list that lack proper motion measurements could also be interesting targets to follow up. It is also possible that the primary neutron star is not a pulsar, and is flying somewhere invisible to us.

\subsection{RX~J1713.7-3946 (G347.3-0.5)}

RX~J1713.7-3946 is one of the closest \acp{SNR} to Earth at an inferred distance of $\sim1$~kpc \citep[]{fuk03}. It has been inferred to be from a type Ib/c SN based on the size of the wind-blown cavity \citep[]{cas04} and the chemical abundance pattern \citep[]{kat15}. Both analyses claim that the progenitor should have originally been a relatively low mass star ($\lesssim20~\msun$) that could not strip its entire hydrogen envelope on its own, implying a binary origin. There was an attempt to discover the remaining binary companion, but so far no optical companion was found in the vicinity of the central compact object 1WGA J1713.4-3949 \citep[]{mig08}. If any stellar companion existed, it has to be of a spectral type later than M or is not a regular main-sequence star.

Although the observational constraints are much weaker, RX~J1713.7-3946 seems to suffer the same problems that Cas A has on the formation of its progenitor. Thus it is possible that the progenitor for RX~J1713.7-3946 was formed through our new scenario too. In fact, this \ac{SNR} is known to have a shell-like morphology with a significantly brighter patch on the North West side \citep[]{koy97,sla99,laz04}, which could be an indication of interaction between the forward shock of the SN and clumpy clouds made of the asymmetric unbound progenitor envelope \citep[e.g.][]{ino12}.

\subsection{G11.2-0.3}

G11.2-0.3 is another core-collapse \ac{SNR} that has been suggested as an evolved version of Cas A \citep[]{gre88}. Analysis of the circumstellar density supports the hypothesis that G11.2-0.3 originated from a stripped-envelope SN \citep[]{bor16}. It again has a shell-like morphology with a bright patch on the South East side \citep[]{tam03,rob03,bor16}. Moreover, the shape of the shell is rather spherical, meaning that the SN itself was not too asymmetric but instead a spherical SN is starting to run into asymmetric circumstellar material. Therefore the morphology of this \ac{SNR} has striking agreements with predictions from our scenario. It may be interesting to search for an optical companion to the central pulsar AX~J1811.5-1926 \citep[]{tor97}. If no strong candidate can be found, the progenitor for this SN could also have been produced through this channel.

\subsection{Other stripped-envelope supernovae}

iPTF13bvn and SN1994I are two other stripped-envelope \acp{SN} that lack companion detections. Both have relatively strong constraints on the photometry of any remaining companion \citep[]{fol16,van16}, placing mass upper limits of $\lesssim10~\msun$. This is deep enough to rule out most evolutionary scenarios involving stable conservative mass transfer \citep[]{ber14,eld15}. It is still possible that the progenitor was stripped through non-conservative mass transfer or common-envelope evolution with a low-mass ($\sim$3--10$~\msun$) stellar companion \citep[]{van16,eld16}. However, the relatively large pre-SN radius for iPTF13bvn ($\sim$30--70$~\rsun$) is difficult to reproduce in the current understanding of common-envelope evolution because the process tightens the orbit too much \citep[]{RH17a}. There are ongoing deeper observations to discover the presence of a low-mass companion, possibly confirming or ruling out the common-envelope channel. If no companion shows up, it leaves us with only two possible scenarios: either the companion is a compact object, or the progenitor was stripped through an ejecta-companion interaction.

\subsection{Type IIL supernovae}

Type II \acp{SN} are usually classified into three subtypes \citep[]{bar79,arc12}. The \acp{SN} that have a plateau in their light curve are classified as type IIP, the ones that show a rapid decline are type IIb, and the intermediate ones type IIL. Type IIP \acp{SN} are the most common kind of \ac{SN} observed, which are known to be explosions of \acp{RSG}. The properties of type IIb \acp{SN} are closer to type Ib SNe, and the progenitors are known to be blue or yellow supergiants from pre-explosion imaging \citep[]{mau04,mau11,dyk14,fol15}. 

The progenitors of type IIL SNe, on the other hand, are much less constrained both observationally and theoretically. \citet[]{and14} pointed out that the distinction between type IIP and IIL SN light curves may be much more ambiguous than previously thought. Nevertheless, type II SN light curves do exhibit a diversity of decline rates, suggesting a wide variety of progenitor properties. Numerical light curve models suggest that the rapid decline of type IIL SNe can be explained if the hydrogen envelope was partially stripped down to $\sim$1--4$~\msun$\footnote{Some studies argue that the envelope mass does not matter but the rapid decline can be explained by different circumstellar matter distributions \citep[]{mor17}. In this case the star does not necessarily need to be stripped.} \citep[e.g.][]{bli93}. If this is the case, the progenitor has to be stripped either through winds or binary interactions. Recent observations find that RSG mass-loss rates are lower than previously considered \citep[]{bea20,hum20}, so self-stripping through winds may not be sufficient to produce type IIL SN progenitors.

Our scenario can produce fully stripped stars, but inevitably produces a similar number of partially stripped stars. For example, our intermediate separation models ($a=$3000--5000$~\rsun$) only strip half of the envelope, with $\sim$4--6$~\msun$ left above the core. For the lower-mass systems the post-interaction $L/M$ ratio does not exceed the threshold value, so is unlikely to develop dynamical envelope instabilities. Therefore the post-interaction mass loss should be relatively low, not leading to a full removal of the envelope. If so, the star could explode with $\sim$1--3$~\msun$ of hydrogen left and appear as a type IIL SN.

\subsection{Rates}
The classification of SNe into spectral types depends on the mass of remaining hydrogen and on the details of the explosion. These in turn are sensitive to the details of the preceding stellar and binary evolution, including the post-interaction evolution of the progenitor star in our scenario, which is beyond the scope of this paper.
Nevertheless, we here attempt to estimate the order-of-magnitude occurrence rate of stripped-envelope SNe through this channel based on simplifying assumptions.

We assume that the post-interaction evolution can remove another $\sim2~\msun$ of the progenitor's hydrogen envelope. Both RSG wind and dynamical pulsations are strongest at the later stages of the RSG phase, where the luminosity is highest. Therefore the timing of the first SN should not affect the total post-interaction mass loss too much. The explosion energy of the first SN is fixed to $E_\mathrm{exp}=10^{51}$~erg. We classify stars with $<1~\msun$ of mass left above the core as stripped-envelope SN progenitors. 

The main requirement of our scenario is that the mass ratio of the binary is sufficiently large that the secondary star will be in its RSG stage by the time the primary explodes. Both the stellar lifetime and duration of the RSG phase depend on the mass of the star, but the dependence is weaker for higher masses (Figure~\ref{fig:ebind_t}). For simplicity, we assume a threshold of $q>0.9$ which is slightly optimistic but sufficient for an order-of-magnitude estimate. Assuming a flat mass ratio distribution \citep[]{san12}, the fraction of systems that have the right mass ratio is $f_q\sim0.1$.

Another requirement is that the binary orbital separation is wide enough so that the secondary \ac{RSG} can fit in its Roche lobe but close enough to lose most of its envelope through ejecta-companion interaction. Based on our simulations, the models with $a\lesssim1.5a_\mathrm{min}$ fulfill the criterion for becoming stripped-envelope SN progenitors. Here $a_\mathrm{min}$ is the separation at which one of the RSGs exactly fills its Roche lobe, which is about $\sim$3 times the stellar radius. By integrating over a log-flat distribution in the range $a\in[0.01\mathrm{AU},1000\mathrm{AU}]$, we find that $f_\mathrm{SE}\sim0.035$ of systems will fall within the range that produces stripped-envelope SN progenitors $a\in[a_\mathrm{min},1.5a_\mathrm{min}]$.

Combining the mass ratio and separation conditions, roughly $f_qf_\mathrm{SE}\sim0.0035$ of the core-collapse population should fall in this category. The observed fraction of stripped-envelope SNe (type Ib/c+IIb) are $\sim35\%$ \citep[]{smi11,li11}, so it implies that 1 out of 100 stripped-envelope SNe could be produced through this channel. Or if we assume that our channel can only create type IIb progenitors, 1 out of 30 type IIb SNe could be explained through this channel.

Here we have not taken into account the detailed mass dependences of the mass ratio threshold, maximum separation, explosion energy fluctuations, etc. We have also chosen a conservative amount of post-interaction mass loss ($\sim2~\msun$), but this should be considerably larger for the relatively higher-mass systems that develop envelope instabilities (see Section~\ref{sec:post-interaction}). We could possibly extend the allowed separation range further for the higher-mass systems. It is also possible that there is a correlation between the period and mass ratio distributions \citep[]{moe17}, and therefore we should not be treating these factors separately. Any other binary configuration is viable as long as it creates a situation where the primary SN occurs when the companion is an RSG. If any such pathways exist in normal binary evolution models, those should be added to the rate.

Type IIL progenitor formation in this scenario more strongly depends on the post-interaction evolution, so we cannot derive any reliable rates. However, because our simulations show that hardly any mass is removed beyond $a>4a_\mathrm{min}$, we can treat that as an upper limit. By integrating over the same orbital separation distribution, the fraction of type IIL candidates formed for a given primary mass becomes $f_\mathrm{IIL}\lesssim0.1$. So the overall rate will be no more than $f_qf_\mathrm{IIL}\lesssim0.01$, which is smaller than the observed type IIL fraction of $\sim6\%$ \citep[]{smi11,and14}.

Our new scenario is restricted to systems that have typically been ignored in binary population synthesis studies. Therefore it does not conflict with any existing binary evolution models but simply adds an extra contribution. The estimated rates we present here seem to be relatively small, so do not affect our overall understanding of stripped-envelope SN progenitor formation. But it may be the dominant or the only channel that can explain the apparently single stripped-envelope SNe like Cas A or iPTF13bvn.

\section{Conclusion}\label{sec:conclusion}
We propose a new scenario for producing stripped-envelope supernova progenitors that do not have binary companions at the time of explosion. The scenario focuses on binary systems in which both components of the binary are red supergiants. The first supernova in the system interacts with the secondary and unbinds most of the secondary envelope. The system is disrupted due to the sudden mass loss and neutron star kick, and the secondary star will be a single stripped-envelope star. Within $\lesssim10^6$~yr, the secondary will explode itself as a stripped-envelope supernova or possibly a type IIL supernova depending on how much mass was stripped in the interaction and its post-interaction evolution.

We investigated the interaction of core-collapse SN ejecta with a \ac{RSG} companion through hydrodynamical simulations. We find that a substantial fraction ($\sim$0.5--0.9) of the envelope can be removed due to the interaction when the orbital separation is within $\lesssim$4--5 times the stellar radius. If the star can lose $\gtrsim1~\msun$ in the post-interaction evolution, it can eventually explode as a stripped-envelope supernova.

We find that the kinematically unbound mass can be well described by the momentum-stripped model proposed by \citet{whe75}, with a modification to take into account the deflection of ejecta. Recombination energy can help to unbind an additional $\sim$1--3$~\msun$; this contribution is larger for higher-mass systems.

After the immediate envelope removal, it is possible that the remaining envelope becomes unstable, especially for higher-mass systems ($M_2\gtrsim19~\msun$). The instability may lead to large pulsations, leading to further substantial mass loss. The spectral type of the second supernova will ultimately depend on how much more mass can be lost in the post-interaction evolution. We leave detailed studies of this phase for future work.

A possible smoking gun observable feature is the shape of the unbound envelope of the secondary that was ejected through the ejecta-companion interaction. More than half of the mass should be ejected in the direction of travel of the supernova ejecta impacting the progenitor. This could form a one-sided shell at $\sim$10--100~pc away from the second supernova site.

We apply our model to the famous supernova remnant Cassiopeia A. Many of the observed properties seem to agree with our scenario including the lack of a companion, the presence of a one-sided envelope shell at $\sim$10--15~pc and a possible runaway pulsar from the first supernova. We also suggest other candidate objects that could have been produced through this channel (RX~J1713.7-3946, G11.2-0.3).

The expected rate of this channel is $\sim$0.35--1$\%$ of the core-collapse supernova population. Therefore we do not expect this to be the major channel for any specific type of supernova. However, it could be largely responsible for the apparently single stripped-envelope supernovae like Cas A or iPTF13bvn.

\section*{Acknowledgements}
The authors thank Satoru Katsuda, Koh Takahashi, Hirotada Okawa, Kotaro Fujisawa, JJ Eldridge, Selma de Mink and Team COMPAS for useful discussions that improved the study. 
RH thanks Lorne Nelson for sharing computational facilities. The computations here were partially carried out on facilities managed by Calcul Qu\'ebec and Compute Canada. 
This work was performed on the OzSTAR national facility at Swinburne University of Technology. The OzSTAR program receives funding in part from the Astronomy National Collaborative Research Infrastructure Strategy (NCRIS) allocation provided by the Australian Government.
TS was supported by the Japan Society for the Promotion of Science (JSPS) KAKENHI grant No. JP19K14739, the Special Postdoctoral Researchers Program, and FY 2019 Incentive Research Projects in RIKEN.
AVG acknowledges funding support by the Danish National Research Foundation (DNRF132).
IM is a recipient of the Australian Research Council Future Fellowship FT190100574.

\section*{Data Availability}
The data underlying this article will be shared on reasonable request to the corresponding author.




\bibliographystyle{mnras}
\bibliography{ref} 



\appendix

\section{Details of the hydrodynamic code}\label{app:code}
The hydrodynamic code HORMONE was originally developed in \citet{RH16}. It is a grid-based code which solves the magneto-hydrodynamic equations through a Godunov-type scheme. The HLLD approximate Riemann solver is used for the numerical flux \citep[]{miy05}, which is equivalent to using the HLLC solver when magnetic fields are neglected. There is a numerical instability known as the carbuncle phenomenon that arise around shocks aligned to the coordinate direction. We remedy this by applying the HLLE solver \citep[]{ein88} around coordinate aligned shocks for fluxes transverse to the shock propagation direction. Shocks are identified using a similar method to \citet{sch16}. For the flux limiter we use the monotonized central limiter \citep[]{lee77} instead of the generalized minmod limiter we were using for previous work. We apply a geometrical correction to the flux limiter that stabilizes the simulation for curvilinear coordinates \citep[]{mig14}.

The main original feature of the code is the ``hyperbolic self-gravity'' solver, which speeds up the self-gravity calculation significantly \citep[]{RH16}. We use a gravitation propagation factor $k_g=10$ \citep[see][]{RH16} and the Robin boundary condition at the outer boundary for the simulations in this paper. The computational domain for the gravitational field is extended $>4$ times beyond the outer boundary for the hydrodynamics to minimize the errors propagating in from the simplified outer boundaries. The computation thus requires more memory, but is still faster than using traditional Poisson solvers.

Spherical coordinates are used for the explosion simulations whereas cylindrical coordinates are used for the ejecta-companion interaction simulations. The cell sizes are increased outwards in a geometrical series in the $r$ direction for spherical coordinates and both $r$ and $z$ directions for the cylindrical coordinate system. The minimum cell sizes $\Delta r_\mathrm{min}$ and $\Delta z_\mathrm{min}$ are chosen to resolve the shortest density scale heights with $>10$ grid points (see Appendix \ref{app:core}). Because we use a regular grid, the sizes of $\Delta r$ and $\Delta z$ can be quite different for cells around the symmetry axis and the $z=0$ plane, resulting in elongated cell shapes. This can potentially be the source of some minor numerical issues, but we did not observe any spurious motions related to the elongation in our simulations.

\section{Numerical treatment of the RSG core}\label{app:core}

Due to the steep pressure gradient and the severe Courant conditions towards the centre of the star, it is common practice to replace the core with a point particle that only interacts with the envelope through gravity. The particle is assumed to have a ``softened'' potential to avoid the singularity at the particle position \citep[e.g.][]{ohl17}. The amount of mass that goes into the point particle $M_\mathrm{pt}$ and the radial extent of the softening $r_\mathrm{s}$ is a somewhat arbitrary choice. Choosing larger values for $M_\mathrm{pt}$ and $r_\mathrm{s}$ reduces the computational cost, but increases the risk of creating spurious artificial errors. For our current purpose, the overall dynamics will not be affected as long as the softening length $r_\mathrm{s}$ is kept much smaller than the stellar radius. There may be some spurious motions as the forward shock traverses the softened region, but this will not affect our overall results as long as the mass contained in this region is small and the bound region is larger than the softened region.

We use a cubic spline form for the softened gravitational potential $\phi'_\mathrm{pt}$ of the point particle \citep[Eq.~(A2) in][]{pri07}. For the gas within the softened region ($r<r_\mathrm{s}$), we construct an artificial density and pressure distribution. There are 4 conditions that the modified density and pressure distributions ($\rho_\mathrm{m}$ and $p_\mathrm{m}$) should satisfy.
\begin{align}
& \nabla p_\mathrm{m}(r)+G\rho_\mathrm{m}(r)\left(\frac{m_\mathrm{m}(r)}{r^2}+M_\mathrm{pt}\nabla\phi'_\mathrm{pt}(r,r_\mathrm{s})\right)=0,\label{eq:hydrostatic}\\
& p_\mathrm{m}(r_\mathrm{s})=p(r_\mathrm{s}),\\
& \nabla p_\mathrm{m}(r_\mathrm{s})=\nabla p(r_\mathrm{s}),\\
& M_\mathrm{pt}+m_\mathrm{m}(r_\mathrm{s})=m(r_\mathrm{s}).\label{eq:mass_constraint}
\end{align}
Here $p(r)$ and $m(r)$ denote the true pressure distribution and mass coordinate of the original star and $m_\mathrm{m}(r)$ is defined as
\begin{equation}
 m_\mathrm{m}(r)\equiv4\pi\int_0^r\rho_\mathrm{m}(r')r'^2dr'.
\end{equation}
The first condition is for hydrostatic equilibrium. Note that the gravitational force from the central point particle is softened, but the self-gravity of the gas obeys Newtonian gravity. The second and third conditions ensure a smooth connection to the pressure distribution outside the softening radius and the fourth condition is required to conserve the total mass. To obtain a unique solution, we further require the entropy of the material in the softened region to be constant. We define entropy as
\begin{equation}
 S\equiv \frac{k_\mathrm{b}}{\mu m_\mathrm{u}}\ln\left({\frac{T^{\frac{3}{2}}}{\rho}}\right)+\frac{4a_\mathrm{rad}T^3}{3\rho},
\end{equation}
where $k_\mathrm{b}$ is the Boltzmann constant, $m_\mathrm{u}$ the atomic mass unit, $a_\mathrm{rad}$ the radiation constant and $\mu$ the mean molecular weight. Temperature $T$ and the mean molecular weight $\mu$ are calculated from the EoS (see Appendix~\ref{app:recombination}).  A uniform entropy distribution helps to avoid developing spurious expansions or convective motions. We solve Equations~(\ref{eq:hydrostatic})--(\ref{eq:mass_constraint}) under the assumption of constant entropy ($S(r)=S(r_\mathrm{s})$) using a shooting method. This gives a unique solution for $\rho_\mathrm{m}(r), p_\mathrm{m}(r)$ and $M_\mathrm{pt}$ given the softening radius $r_\mathrm{s}$.

After replacing the central part with this artificial density and pressure distribution, the steepest density gradient that needs to be resolved is roughly the gradient at the softening length $\nabla\rho(r_\mathrm{s})$. We make our minimum gridsize 10 times smaller than this minimum density scale height
\begin{equation}
 \Delta r_\mathrm{min}=\Delta z_\mathrm{min}=\frac{1}{10}\frac{\rho(r_\mathrm{s})}{\rho'(r_\mathrm{s})}.
\end{equation}

We choose a softening radius of $r_\mathrm{s}=22~(28)~\rsun$ for our 16~(19)$~\msun$ RSG model. With this procedure we have tested that the core does not develop any artificial motions at least for several dynamical timescales. We have tested on spherical and cylindrical coordinates and it works in both cases. 

\section{Numerical treatment of recombination energy}\label{app:recombination}

Similar to some other hydrodynamical simulations that incorporate recombination energy, we account for recombination energy by changing the \ac{EoS}. Previous studies have implemented this by using tabulated EoSs generated from more detailed codes \citep[e.g.][]{nan15}. Here we use a slightly different approach which is conceptually similar but more memory-efficient (and possibly computationally cheaper). 

The physical quantities that are directly used in the hydrodynamics are density $\rho$, momentum $\rho\mathbfit{v}$, and total energy $e$. Specific internal energy is computed by subtracting kinetic energy from the total energy $\varepsilon_\mathrm{int}=e/\rho-\frac{1}{2}\mathbfit{v}^2$. For EoS A, the pressure simply follows from $p=(\gamma-1)\rho\varepsilon_\mathrm{int}$, where $\gamma=5/3$ is the adiabatic index. For the other EoSs, we first compute the temperature $T$ by finding the root of 
\begin{equation}
 \varepsilon_\mathrm{int}=\frac{3}{2}\frac{k_\mathrm{b}T}{\mu(x_i)m_\mathrm{u}}+\frac{a_\mathrm{rad}T^4}{\rho}+\varepsilon_\mathrm{ion}(x_i),\label{eq:energyequation}
\end{equation}
and then the pressure can be computed from
\begin{equation}
 p = \frac{\rho k_\mathrm{b}T}{\mu(x_i)m_\mathrm{u}}+\frac{a_\mathrm{rad}T^4}{3}.\label{eq:EOS}
\end{equation}
The mean molecular weight $\mu$ and ionization energy $\varepsilon_\mathrm{ion}$ are functions of ionization fractions $x_i=x_i(\rho,T)$, where the index $i$ runs over the number of ionization states. EoS B does not include the last term in Eq.~(\ref{eq:energyequation}) and uses a fixed mean molecular weight, assuming that all the gas is always fully ionized. With the additional ionization term, part of the recombination energy can be stored in the third term until the ionization fraction changes and is released as thermal energy. Because our equations are adiabatic and do not include any radiation transport, all of the released energy will be fully thermalized. This assumption is valid for optically thick regions where radiative losses can be neglected. For the simulations presented in this paper, the regions where recombination occurs are mostly optically thick, justifying our adiabatic treatment.

Ionization fractions $x_i$ can in principle be obtained by solving the Saha equations. However, solving the Saha equations and Eq.~(\ref{eq:energyequation}) self-consistently is computationally demanding, given that the EoS module is called heavily throughout the simulation. Instead we perform analytical fits to the MESA EoS to obtain expressions for $x_i$. There are four main ionizations included in the MESA EoS, which are for molecular hydrogen, ionization of hydrogen, single and double ionizations of helium. Our fitting formulae are
\begin{align}
& \varepsilon_\mathrm{ion}=\sum_{i=1}^4\varepsilon_ix_i,\label{eq:fit_energy}\\
& \mu=4\left[2(x_1+2x_2)X+(x_3+x_4-1)Y+2\right]^{-1},\label{eq:mean_molecular_weight}\\
& x_i=\frac{1}{2}\left[\tanh\left(\frac{\log T-T_i(\log Q)}{\sigma_i(\log Q)}\right)+1\right],\label{eq:ionization}\\
& T_i(\theta)=a_i(1-fY)\log\left(\varepsilon_i\frac{m_\mathrm{u}}{k_b}\right)+b_i\theta,\label{eq:ionization_temperature}\\
& \sigma_i(\theta)=c_iT_i(\theta)(1+d_i\theta),\label{eq:fit_width}
\end{align}
where $\log Q\equiv\log\rho-2\log T+12$ in cgs units and $Y$ is the helium mass fraction. $\varepsilon_i$ are the ionization energies for a given mixture of gas $\varepsilon_1=\frac{1}{2}X\mathcal{X}_\mathrm{H_2}, \varepsilon_2=X\mathcal{X}_\mathrm{H^+}, \varepsilon_3=\frac{1}{4}Y\mathcal{X}_\mathrm{He^+}, \varepsilon_4=\frac{1}{4}Y\mathcal{X}_\mathrm{He^{++}}$, where $X$ and $Y$ are the hydrogen and helium mass fractions. The ionization energies for each element are $\mathcal{X}_\mathrm{H_2}=4.36\times10^{12}$~erg~mol$^{-1}, \mathcal{X}_\mathrm{H^+}=1.312\times10^{13}$~erg~mol$^{-1}, \mathcal{X}_\mathrm{He^{+}}=2.3723\times10^{13}$~erg~mol$^{-1}, \mathcal{X}_\mathrm{He^{++}}=5.2505\times10^{13}$~erg~mol$^{-1}$. $T_i$ corresponds to the common logarithm of the ionization temperature and $\sigma_i$ the width of the transition in log temperature space. Values of the fitting coefficients $a_i, b_i, c_i, d_i, f$ are given in Table \ref{tab:recfits}. 

The hyperbolic tangent function in Eq.~(\ref{eq:ionization}) can be computationally expensive, so we approximate it by 
\begin{equation}
 \tanh(x)\sim\frac{x^5+105x^3+945x}{15(x^4+28x^2+63)},
\end{equation}
and we clip the function at $|x|\sim3.647$ so that the function does not go beyond positive or negative unity. The approximation closely resembles the tanh function so does not change our results. When factorized appropriately, this only requires 10 primitive operations (5 multiplications, 1 division and 4 summations) so is much more computationally efficient than evaluating $\tanh$ functions. 

\begin{table}
 \begin{center}
  \caption{Fitting coefficients for the ionization fractions. Note that $a_1$ and $b_1$ change their values at a density $\rho_t=4\times10^{-10}$~g~cm$^{-3}$.\label{tab:recfits}}
  \begin{tabular}{ccccc}
   \hline
   &$i=1$&$i=2$&$i=3$&$i=4$\\\hline
   $a_i$&0.751 ($\rho<\rho_t$)&0.821&0.829&0.846\\
   &0.753 ($\rho\geq\rho_t$)&&&\\
   $b_i$&0.000 ($\rho<\rho_t$)&0.055&0.055&0.055\\
   &0.055 ($\rho\geq\rho_t$)&&&\\
   $c_i$&0.02&0.025&0.015&0.015\\
   $d_i$&0.05&0.05&0.05&0.05\\
   $f$&&0.005&&\\\hline   
  \end{tabular}
 \end{center}
\end{table}

\begin{figure}
 \centering
 \includegraphics[]{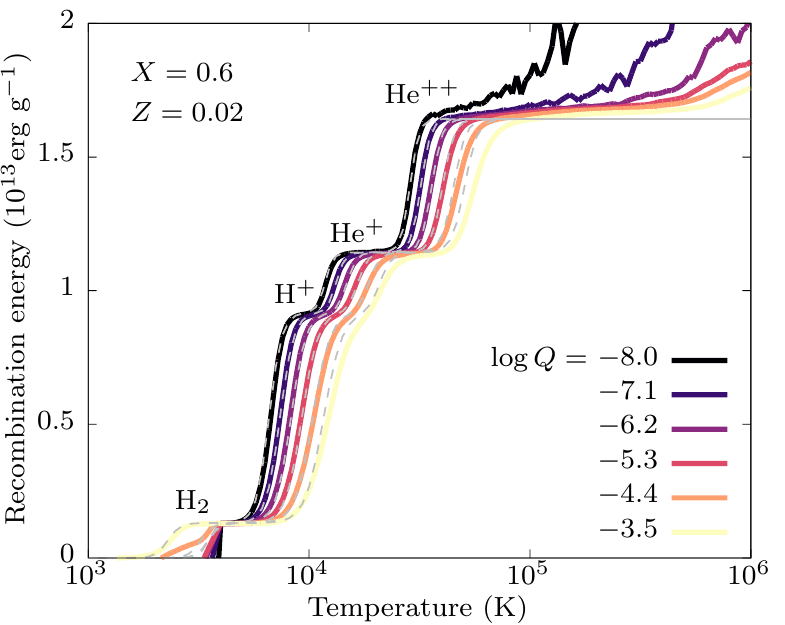}
 \caption{Recombination energy in the MESA EoS as a function of temperature for various values of $\log Q$. The chemical composition is fixed to $X=0.6$ and $Z=0.02$. Grey dashed curves are the fitting functions given in Eqs.\ref{eq:fit_energy}--\ref{eq:fit_width}. Although the fits are not colour-coded, it should be clear which curve they are intended to fit. Species responsible for each transition are labelled accordingly.\label{fig:erec_fit}}
\end{figure}

We compare our fitting functions to the MESA EoS in Figure~\ref{fig:erec_fit}. The recombination energy is extracted from the MESA EoS by subtracting the contribution of ideal gas and radiation from the internal energy given in the EoS table by
\begin{equation}
 \varepsilon_\mathrm{ion,MESA}=\varepsilon_\mathrm{MESA}-\frac{3P_\mathrm{gas}}{2\rho}-\frac{a_\mathrm{rad}T^4}{\rho}.
\end{equation}
Here $\varepsilon_\mathrm{MESA}, P_\mathrm{gas}$ are the specific internal energy and gas pressure given in the EoS table, respectively. The four major transitions are fit remarkably well for the range $\log Q\lesssim-5$. The MESA EoS curves show irregular features in the higher-temperature regions ($T>3\times10^4$K) that are likely contributions from heavier elements, but we do not attempt to model this part. In the higher-$\log Q$ range, the He recombination temperatures are slightly underestimated in the fitted functions. 

\begin{figure}
 \centering
 \includegraphics[]{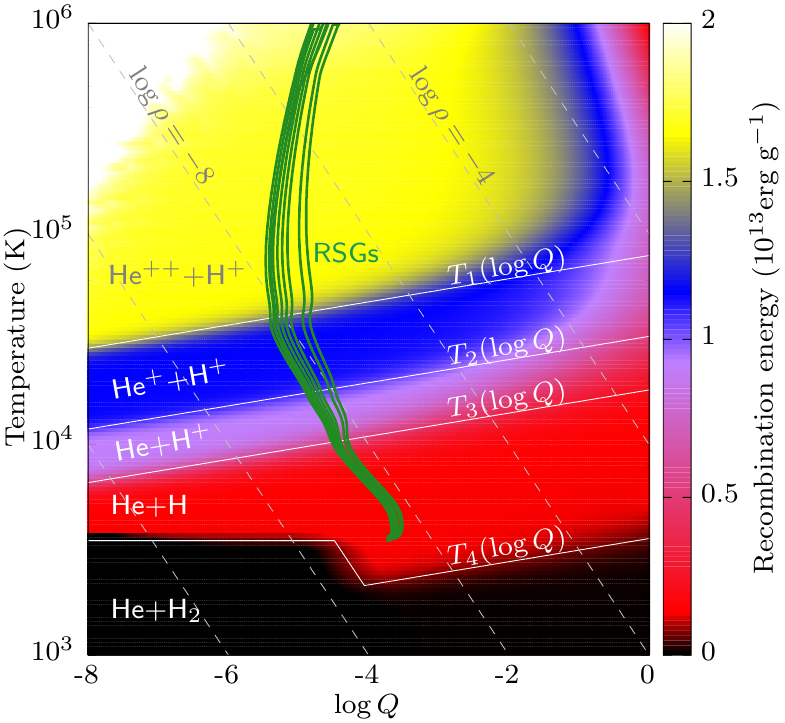}
 \caption{2D landscape of recombination energy in the MESA EoS as a function of $\log Q$ and temperature with a fixed chemical composition $X=0.6, Z=0.02$. Grey dashed lines are drawn along fixed densities. White lines show our fits to the ionization temperatures given by Eq.~(\ref{eq:ionization_temperature}) with the fitting coefficients in Table \ref{tab:recfits}. Green curves are various RSG models with masses in the range $12\leq M/\msun\leq20$.\label{fig:erec_landscape}}
\end{figure}

This can be better observed in Figure~\ref{fig:erec_landscape} where we are looking at Figure~\ref{fig:erec_fit} from above. Each coloured region corresponds to the plateaus seen in Figure~\ref{fig:erec_fit}. The transition temperatures given in Eq.~(\ref{eq:ionization_temperature}) were fitted to the boundaries between the different regions. The boundaries start bending upwards beyond $\log Q\gtrsim-4$, meaning our fits break down. However, this region in the $Q$--$T$ plane is never reached in our simulations. We show various RSG models in the range $12\leq M/\msun\leq20~\msun$ as green curves. They only occupy a very narrow range of the $Q$--$T$ plane. The simulation domain covers a much wider region, but only in the lower-$Q$ direction and not towards higher $Q$. For very low density cases ($\rho<10^{-14}$~g~cm$^{-3}$) we simply use the transition temperatures at $\rho=10^{-14}$~g~cm$^{-3}$. The fit was also carried out for the table with different compositions ($X=0.4, 0.6, 0.8$ and $Z=0, 0.01, 0.02$). The basic shape can be fit with the same function and we adjusted minor deviations through the factor $f$. The matter in our simulation has a composition of $X\sim0.72, Y\sim0.27$, which lies between the various tables we have fit our functions to. 

With these fits to the ionization energy and mean molecular weight, Eq.~(\ref{eq:energyequation}) becomes a single variable root finding problem. Since our expression for the ionization energy has many inflection points in the function, traditional root finding algorithms such as the Newton-Raphson method do not converge unless the initial guess is already close to the solution. Here we overcome this problem by applying the novel W4 method \citep[]{oka18,fuj19} which is an iterative root finding algorithm for nonlinear systems of equations that can find solutions starting from arbitrary guesses provided the root function is smooth and continuous. For single variable problems the W4 method can be simplified as
\begin{equation}
 \begin{cases}
 x_{n+1}=x_n+\Delta\tau p_n\\
 p_{n+1}=(1-2\Delta\tau)p_n-\Delta\tau \frac{f(x_n)}{f'(x_n)}
\end{cases}
\end{equation}
where $f(x)=0$ is the equation to be solved, $f'(x)$ is the derivative and $p_n$ is a variable that is unrelated to $f(x)$. It is known that the method converges as long as the virtual time step is taken as $0<\Delta\tau<1$. However, the method sacrifices computational time to achieve global convergence, and the Newton-Raphson method is much faster when the initial guess is close enough to the solution. Thus we employ a hybrid method where we use the W4 method until it reaches the vicinity of the solution ($|f(x)/(xf'(x))|<0.01$) and then switch to the Newton-Raphson method until it converges ($|f(x)/(xf'(x))|<10^{-10}$). This allows us to make use of the fast convergence of the Newton-Raphson method while acquiring the global convergence of the W4 method at the same time.

Since our method utilizes an analytical fit to a tabulated \ac{EoS}, the results are basically the same as directly using the EoS table. However, our method has several advantages over tabular EoSs. For example, it is known that when using EoS tables, the interpolation scheme has a large influence on the thermodynamic consistency \citep[]{swe96}. With crude interpolation schemes, this could lead to unphysical entropy build-up that could cause trouble in the long term. Since our EoS is expressed in an analytical functional form, our method should naturally satisfy thermodynamic consistency with an accuracy of the tolerance applied for the root finding. Our method is also easy to customize. Various ionization energies can be added or subtracted in a simple way. For example, we set $\varepsilon_1=\varepsilon_2=0$ when we want to effectively account for inefficient thermalisation of hydrogen recombination energy in optically thin regions (EoS C1). This still takes into account the change of mean molecular weight due to the recombination through Eq.~(\ref{eq:mean_molecular_weight}).


\bsp	
\label{lastpage}
\end{document}